\PassOptionsToPackage{unicode}{hyperref}
\PassOptionsToPackage{hyphens}{url}
\documentclass[
]{article}
\usepackage{xcolor}
\usepackage{amsmath,amssymb}
\usepackage{iftex}
\ifPDFTeX
  \usepackage[T1]{fontenc}
  \usepackage[utf8]{inputenc}
  \usepackage{textcomp} 
\else 
  \usepackage{unicode-math} 
  \defaultfontfeatures{Scale=MatchLowercase}
  \defaultfontfeatures[\rmfamily]{Ligatures=TeX,Scale=1}
\fi
\usepackage{lmodern}
\ifPDFTeX\else
\fi
\IfFileExists{upquote.sty}{\usepackage{upquote}}{}
\IfFileExists{microtype.sty}{
  \usepackage[]{microtype}
  \UseMicrotypeSet[protrusion]{basicmath} 
}{}
\makeatletter
\@ifundefined{KOMAClassName}{
  \IfFileExists{parskip.sty}{%
    \usepackage{parskip}
  }{
    \setlength{\parindent}{0pt}
    \setlength{\parskip}{6pt plus 2pt minus 1pt}}
}{
  \KOMAoptions{parskip=half}}
\makeatother
\usepackage{color}
\usepackage{fancyvrb}

\DefineVerbatimEnvironment{Highlighting}{Verbatim}{commandchars=\\\{\}}
\newenvironment{Shaded}{}{}

\newcommand{\NormalTok}[1]{#1}

\usepackage{longtable,booktabs,array}
\usepackage{calc} 
\usepackage{etoolbox}
\makeatletter
\patchcmd\longtable{\par}{\if@noskipsec\mbox{}\fi\par}{}{}
\makeatother
\IfFileExists{footnotehyper.sty}{\usepackage{footnotehyper}}{\usepackage{footnote}}
\makesavenoteenv{longtable}
\providecommand{\tightlist}{%
  \setlength{\itemsep}{0pt}\setlength{\parskip}{0pt}}
\usepackage[]{natbib}
\usepackage{graphicx}
\usepackage{amsmath,amssymb}
\usepackage{newunicodechar}
\newunicodechar{→}{\ensuremath{\rightarrow}}
\newunicodechar{↔}{\ensuremath{\leftrightarrow}}
\newunicodechar{≥}{\ensuremath{\geq}}
\newunicodechar{≤}{\ensuremath{\leq}}
\newunicodechar{−}{\ensuremath{-}}
\newunicodechar{∈}{\ensuremath{\in}}
\newunicodechar{∉}{\ensuremath{\notin}}
\newunicodechar{⊆}{\ensuremath{\subseteq}}
\newunicodechar{≠}{\ensuremath{\neq}}
\newunicodechar{≈}{\ensuremath{\approx}}
\newunicodechar{∅}{\ensuremath{\emptyset}}
\newunicodechar{÷}{\ensuremath{\div}}
\newunicodechar{∎}{\ensuremath{\blacksquare}}
\newunicodechar{…}{\dots}
\newunicodechar{§}{\S}
\providecommand{\real}[1]{#1}
\providecommand{\tightlist}{\setlength{\itemsep}{0pt}\setlength{\parskip}{0pt}}
\usepackage{bookmark}
\IfFileExists{xurl.sty}{\usepackage{xurl}}{} 
\hypersetup{
  pdftitle={A Deterministic Control Plane for LLM Coding Agents},
  pdfauthor={Padmaraj Madatha \textbackslash{} Happiest Minds Technologies (AIP Centre of Excellence)},
  hidelinks,
  pdfcreator={LaTeX via pandoc}}

\title{A Deterministic Control Plane for LLM Coding Agents}
\author{Padmaraj Madatha \textbackslash{} Happiest Minds Technologies
(AIP Centre of Excellence)}
\date{June 2026}

\begin{document}
\maketitle
\begin{abstract}
LLM coding harnesses grant agents broad file and shell access, yet the
configuration layer that steers them-rules files, agent definitions,
IDE-specific markdown-is largely unmanaged. A prevalence study of 10,008
public GitHub repositories (\emph{n}=6,145 agent config files)
establishes one robust headline finding: agent configurations
\textbf{propagate as undeclared shared components}-10.1\% of tracked
paths are exact duplicates across independent repositories
(fork-adjusted; measured by SHA-256, threshold-independent), and 75.5\%
of clone pairs cross organisational boundaries. Two further patterns are
\emph{indicative} rather than definitive: configurations are
\textbf{rarely revised} (a 58\% single-commit majority; the gap narrows
but persists after age-normalisation-0.4 versus 0.6 commits/month
against CI/CD workflows in the same repos), and they \textbf{rarely
declare permission boundaries} (\textless1\% of agent configs versus
33\% of Actions workflows, on a fragile parser with \emph{n}=31 true
positives). A fourth gap-\textbf{unbounded execution} without enforced
traceability-is not observable from static config files; a
fifth-\textbf{expertise locked to harness dialects}-is only partially
visible in the corpus but is structurally evident from incompatible
config formats across tools.

We propose a \textbf{deterministic control plane} above the harness (not
replacing it) that maps one-to-one to these gaps. \textbf{Rel(AI)Build},
our reference implementation, treats agent definitions as a managed
supply chain (SHA-256 content addressing, HMAC-stamped lockfiles,
hash-chained audit logs) to counter undeclared propagation; enforces
tiered permissions and attack-derived blocklists before LLM invocation
where configs declare none; gates feature work through a phase state
machine with requirement→file→test traceability to bound execution;
compiles a single canonical definition to seven IDE targets to port
expertise across dialects; and detects prompt drift via Jaccard
similarity. Conformance tests on injected violations confirm each
mechanism enforces its stated invariant; developer outcomes remain
future work. \textbf{Determinism applies to install-time gates,
blocklists, and phase-state transitions; trace linkage is cooperative
and auditable post-hoc, not pre-execution enforced.} Governance of this
layer must be \textbf{deterministic and tool-agnostic}-not delegated to
further LLM orchestration.
\end{abstract}

\subsection{1. Introduction}\label{introduction}

\subsubsection{1.1 The thin governance
surface}\label{the-thin-governance-surface}

Modern LLM coding harnesses combine a capable execution loop - codebase
indexing, semantic retrieval, MCP tools, and multi-turn file/shell
actions - with a comparatively thin \textbf{governance surface}:
organisations typically express policy through natural-language
configuration files (\texttt{CLAUDE.md}, \texttt{.cursor/rules}, Copilot
instructions, \texttt{AGENTS.md}) that are injected into context
alongside retrieved code and task input. That low-friction authoring
model drove adoption, but it leaves agent definitions largely unmanaged
at enterprise scale.

However, that thin governance surface becomes a liability in regulated,
multi-developer, multi-repository organisations - even as harness
runtimes grow more capable at retrieving and acting on code. Three
structural gaps recur:

\begin{enumerate}
\def\labelenumi{\arabic{enumi}.}
\item
  \textbf{The configuration is unmanaged software.} A
  \texttt{.cursorrules} file that encodes hard-won architectural
  conventions is copy-pasted between repositories, edited
  inconsistently, and trusted without provenance. Codebase indexing and
  semantic retrieval improve \emph{what the agent knows about the
  repository}, but they provide no supply-chain guarantees for the rules
  files themselves. There is no notion of \emph{this is the approved
  version}, \emph{this file has not been tampered with}, or \emph{who
  changed it and when}. Yet this file directly steers an agent with
  write and execute access. Recent literature demonstrates that LLMs
  frequently hallucinate software packages, and attackers actively
  weaponise these hallucinations via typosquatting
  \citep{vulcan, lanyado}. The configuration that shapes an agent's
  dependency suggestions requires at least the same supply-chain rigour
  currently applied to standard software dependencies.
\item
  \textbf{Execution is unbounded.} Bare harnesses impose no process.
  There is no enforced requirement→implementation→test trace, no
  mandatory review checkpoint, and - critically - no hard cap on
  self-correction loops. Recent empirical work on LLM code debugging
  shows effectiveness decays sharply within 2--3 attempts \citep{ddi};
  standards for generative and agentic AI emphasise proportional human
  oversight, bounded autonomy, and escalation when systems act on a
  user's behalf \citep{nist-gai, owaspllm, berkeley-agentic}.
\item
  \textbf{Expertise does not port.} Each harness has its own
  configuration dialect, tool-permission syntax, and file layout (Table
  2). An organisation that standardises on more than one tool - common,
  since different teams prefer different IDEs - must re-author and
  re-maintain the same expertise multiple times.
\end{enumerate}

A prevalence study (§7) quantifies the observable proxies for Gaps 1 and
3 in public repositories (Table 1b); Gap 2 is a process-layer risk that
static config analysis cannot measure and motivates §4.5 independently.
The robust headline finding is widespread exact-duplicate propagation
(\textbf{10.1\%} fork-adjusted; 75.5\% cross-organisation); two
indicative proxies - shallow version-control depth (58\% single-commit)
and near-absence of declared permission boundaries (\textless1\% of
agent config files) - point the same direction but are weaker
(age-confounded and parser-fragile respectively). Full statistics,
confidence intervals, and age-normalised rates are in §7.2 and Figures
8--9.

\begin{center}\includegraphics[width=\linewidth,height=0.6\textheight,keepaspectratio]{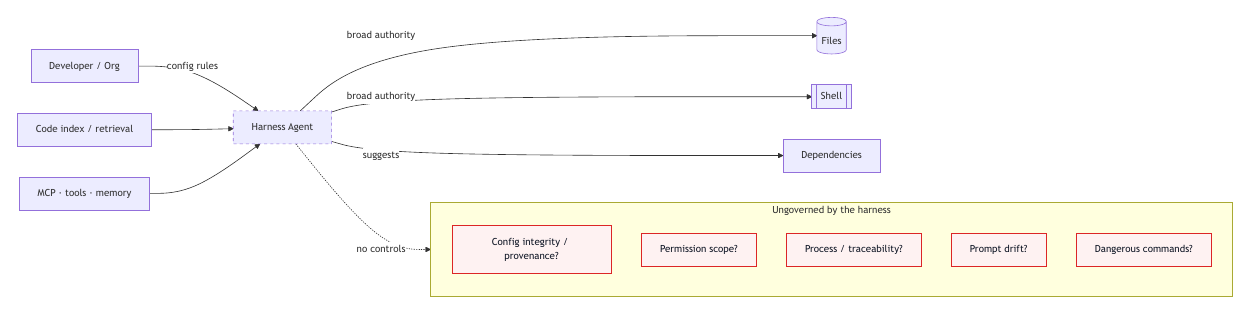}\end{center}

\emph{Figure 1. The thin-governance gap. Runtime augmentation (indexing,
tools, MCP) improves execution, but the harness provides no governance
over five risk surfaces (red). These ungoverned risks sit adjacent to
the harness - present but unaddressed - motivating the deterministic
control plane in §3.}

\subsubsection{1.2 Thesis}\label{thesis}

Our thesis is deliberately narrow. Role-based access control (RBAC),
sandboxing, and orchestration are established paradigms; we do not claim
them as novel. Instead:

\begin{quote}
\textbf{The configuration-and-process layer that surrounds an LLM coding
agent can be treated as a managed software supply chain, governed
\emph{deterministically} and \emph{independently of the underlying
harness or model}, addressing critical integrity, traceability, and
portability risks that the harness alone does not.}
\end{quote}

The word \emph{deterministic} is load-bearing. Every governance
mechanism we propose is implemented in ordinary, testable code (hashing,
regex matching, a state machine, set arithmetic), not delegated to an
LLM. A non-deterministic component cannot serve as a trustworthy control
for another non-deterministic component.

\subsubsection{1.3 Contributions}\label{contributions}

We propose a control plane architecture and detail its reference
implementation, offering five architectural contributions:

\begin{enumerate}
\def\labelenumi{\arabic{enumi}.}
\tightlist
\item
  \textbf{A supply-chain architecture for agent definitions} (§4.1, §5):
  content addressing (SHA-256), HMAC-stamped lockfiles for
  accidental/automated corruption detection, append-only hash-chained
  audit logs, install-time integrity gates, and an agent-configuration
  SBOM mapping to NTIA minimum elements \citep{ntia} (EO 14028
  \citep{eo14028}, SLSA Build Level 2 \citep{slsa}).
\item
  \textbf{A deterministic pre-execution guardrail system} (§4.2, §4.3):
  tier-based permissions bound to the agent definition's supply chain
  and enforced prior to LLM invocation, alongside an attack-derived
  command/path blocklist deployed as IDE runtime hooks (Claude Code
  \texttt{PreToolUse}, Cursor \texttt{beforeShellExecution}) and a
  post-delegation \texttt{scan-diff} gate (§4.5).
\item
  \textbf{A tokenisation-normalisation pipeline and drift detection
  mechanism} (§4.4) measuring cosmetic prompt edits as an operational
  risk surface using Jaccard similarity \citep{sclar} - a detection
  mechanism, not a validated behavioural predictor (§9.1).
\item
  \textbf{A deterministic phase-gated lifecycle} (§4.5) whose state
  machine enforces process invariants (ordering, mandatory delegation,
  requirement→file→test traceability) and blocks execution upon
  violation.
\item
  \textbf{A define-once / compile-to-many transformer} (§4.6) portable
  across seven IDE targets, with a governance-preservation argument
  (Appendix B).
\end{enumerate}

\textbf{Determinism scope (not a contribution):} Pre-execution controls
(install gates, blocklists, phase-state transitions) are deterministic;
requirement→file→test linkage depends on cooperative agent invocation of
\texttt{trace-update} and is audited post-hoc, not LLM-enforced (§4.5
trust boundary).

Empirical activities - implementation conformance testing (§6),
prevalence measurement (§7), and a DRIFT-detection vignette (Appendix D)
- are reported separately and are not listed as architectural
contributions.

\textbf{Table 1. Evidence mapping (what each section establishes).}

{\def\LTcaptype{none} 
\begin{longtable}[]{@{}
  >{\raggedright\arraybackslash}p{(\linewidth - 4\tabcolsep) * \real{0.3333}}
  >{\raggedright\arraybackslash}p{(\linewidth - 4\tabcolsep) * \real{0.3333}}
  >{\raggedright\arraybackslash}p{(\linewidth - 4\tabcolsep) * \real{0.3333}}@{}}
\toprule\noalign{}
\begin{minipage}[b]{\linewidth}\raggedright
Claim
\end{minipage} & \begin{minipage}[b]{\linewidth}\raggedright
Evidence today
\end{minipage} & \begin{minipage}[b]{\linewidth}\raggedright
Not claimed
\end{minipage} \\
\midrule\noalign{}
\endhead
\bottomrule\noalign{}
\endlastfoot
Supply chain, guardrails, drift, phases, compile & §6 conformance tests
(injected violations, N=10/15/20) & Outcome superiority \\
Governance gap exists & §7 prevalence (duplicates, depth, age-normalised
rates) & Rel(AI)Build deployment impact \\
Mechanism illustration & Appendix D DRIFT vignette (N=1, disclaimers) &
Comparative effectiveness \\
\end{longtable}
}

\textbf{Table 1b. Structural gaps (§1.1) → prevalence proxies (§7).}

{\def\LTcaptype{none} 
\begin{longtable}[]{@{}
  >{\raggedright\arraybackslash}p{(\linewidth - 6\tabcolsep) * \real{0.2500}}
  >{\raggedright\arraybackslash}p{(\linewidth - 6\tabcolsep) * \real{0.2500}}
  >{\raggedright\arraybackslash}p{(\linewidth - 6\tabcolsep) * \real{0.2500}}
  >{\raggedright\arraybackslash}p{(\linewidth - 6\tabcolsep) * \real{0.2500}}@{}}
\toprule\noalign{}
\begin{minipage}[b]{\linewidth}\raggedright
Structural gap
\end{minipage} & \begin{minipage}[b]{\linewidth}\raggedright
Observable in corpus?
\end{minipage} & \begin{minipage}[b]{\linewidth}\raggedright
Proxy metric
\end{minipage} & \begin{minipage}[b]{\linewidth}\raggedright
§7 finding
\end{minipage} \\
\midrule\noalign{}
\endhead
\bottomrule\noalign{}
\endlastfoot
\textbf{Gap 1:} Config is unmanaged software & Yes & Exact-duplicate
propagation; commit depth; credential patterns & 10.1\% fork-adjusted
duplicates; 58\% single-commit files; 3.18\% credential hits \\
\textbf{Gap 2:} Execution is unbounded & No (process not in config
files) & - & \emph{Architectural motivation only; not measured by §7} \\
\textbf{Gap 3:} Expertise does not port & Partially & Multi-dialect
tracked paths; permission-syntax absence & \textless1\% permission
declarations; six config dialects in corpus \\
\end{longtable}
}

The evidence reported is preliminary, characteristic of early-stage SE
research contributions; a controlled developer study (§9.2) is required
before claiming productivity outcomes.

\begin{center}\rule{0.5\linewidth}{0.5pt}\end{center}

\subsection{2. Background and Related
Work}\label{background-and-related-work}

The landscape of AI coding tools has evolved rapidly. Our architecture
sits within this ecosystem but serves a distinct, complementary purpose.

\subsubsection{2.1 LLM coding harnesses}\label{llm-coding-harnesses}

Cursor, Claude Code, GitHub Copilot, OpenAI Codex CLI, and Windsurf
provide editor- or CLI-integrated agents with file and shell tools,
steered by per-project configurations. They differ in native context
augmentation, context scoping, and dialect (Table 2) but share the thin
\textbf{governance} interface of §1.1. Projects such as Aider
\citep{aider} follow a similar per-project configuration pattern. They
are the \emph{substrate} our architecture runs on; we do not reimplement
their model inference, codebase indexing, or native editing loops.

\textbf{Table 2. Configuration surface of common harnesses (approximate;
formats evolve rapidly).}

{\def\LTcaptype{none} 
\begin{longtable}[]{@{}
  >{\raggedright\arraybackslash}p{(\linewidth - 10\tabcolsep) * \real{0.1667}}
  >{\raggedright\arraybackslash}p{(\linewidth - 10\tabcolsep) * \real{0.1667}}
  >{\raggedright\arraybackslash}p{(\linewidth - 10\tabcolsep) * \real{0.1667}}
  >{\raggedright\arraybackslash}p{(\linewidth - 10\tabcolsep) * \real{0.1667}}
  >{\raggedright\arraybackslash}p{(\linewidth - 10\tabcolsep) * \real{0.1667}}
  >{\raggedright\arraybackslash}p{(\linewidth - 10\tabcolsep) * \real{0.1667}}@{}}
\toprule\noalign{}
\begin{minipage}[b]{\linewidth}\raggedright
Harness
\end{minipage} & \begin{minipage}[b]{\linewidth}\raggedright
Config file(s)
\end{minipage} & \begin{minipage}[b]{\linewidth}\raggedright
Tool-permission syntax
\end{minipage} & \begin{minipage}[b]{\linewidth}\raggedright
Context scoping
\end{minipage} & \begin{minipage}[b]{\linewidth}\raggedright
Native context augmentation
\end{minipage} & \begin{minipage}[b]{\linewidth}\raggedright
Runtime enforcement
\end{minipage} \\
\midrule\noalign{}
\endhead
\bottomrule\noalign{}
\endlastfoot
Cursor & \texttt{.cursor/rules/*.mdc} & implicit & glob / file &
Codebase index, \texttt{@} retrieval, rules globs, MCP & Implicit
scoping + \texttt{beforeShellExecution} hook \\
Claude Code & \texttt{CLAUDE.md}, \texttt{.claude/agents/*} & comma
string & project & Project memory, subagents, MCP, tool hooks &
Session-start tool allowlist + \texttt{PreToolUse} hooks \\
Copilot (VS Code) & \texttt{*.agent.md}, instructions & YAML array &
file / PR & Workspace indexing, agent tools, instructions & None at
config layer \\
Aider & \texttt{.aider.conf.yml}, conventions & - & project & Repo map /
git context & Shell access; no config-layer enforcement \\
Codex & root \texttt{AGENTS.md} & - & file & File-scoped context &
None \\
Windsurf & \texttt{.windsurfrules} & - & file & Repo-aware Cascade
context & None \\
\end{longtable}
}

\subsubsection{2.2 Orchestration frameworks and Autonomous
Agents}\label{orchestration-frameworks-and-autonomous-agents}

Frameworks like \textbf{LangGraph}, \textbf{CrewAI}, and
\textbf{AutoGen} \citep{autogen} orchestrate multiple LLM ``roles''
through complex graphs and conversation patterns. Similarly,
\textbf{Devin}, \textbf{SWE-Agent} \citep{sweagent}, and
\textbf{OpenHands} \citep{opendevin} (formerly OpenDevin) provide
standalone, autonomous environments focused on resolving software issues
end-to-end. Benchmark harnesses such as AgentBench \citep{agentbench}
provide standardised evaluation environments for autonomous agent
capabilities.

Our architecture differs fundamentally in its goal. We are not proposing
a new graph-based communication protocol for agents, nor are we building
a standalone autonomous sandbox. Instead, we propose a \emph{governance
overlay} for existing IDE-integrated harnesses. Since 2023, adoption has
shifted toward editor-integrated agents (Cursor, Claude Code, Copilot)
with per-project configuration files rather than standalone multi-agent
sandboxes; our architecture targets that deployment model. Where
frameworks like CrewAI focus on the \emph{dynamic coordination} of
agents, our architecture focuses on the \emph{deterministic supply-chain
security, integrity, and traceability} of the prompts and permissions
that define those agents.

\subsubsection{2.3 Agent Observability and
Governance}\label{agent-observability-and-governance}

Tools such as \textbf{AgentOps} and \textbf{OpenTelemetry} integrations
for LLMs provide essential \emph{runtime observability} (tracking token
counts, tracing API calls, logging errors). \textbf{Guardrails AI}
\citep{guardrails} and \textbf{NeMo Guardrails} \citep{nemoguardrails}
(NVIDIA) validate LLM \emph{outputs} at inference time - complementary
to, but distinct from, our focus on the \emph{configuration-and-process}
layer. Amershi et al. \citep{amershi} establish software-engineering
process governance as a first-class concern for ML systems; we apply the
same principle upstream to agent definitions. Our architecture focuses
on \emph{preventative, pre-execution governance}: install-time
permission checking and state-machine gates that \emph{block}
out-of-policy configurations before the LLM can invoke them.

Model providers publish governance frameworks at the capability and
deployment layer - Anthropic's Responsible Scaling Policy and agent
safety levels \citep{anthropic-rsp}, OpenAI's Preparedness Framework
\citep{openai-prep} - which tier model capabilities and deployment gates
rather than governing canonical agent-definition supply chains or phase
traceability. Recent empirical work on IDE-integrated coding agents
characterises the \emph{runtime} prompt-injection attack surface: Liu et
al. \citep{liu-aishell} report high attack-success rates on Cursor and
GitHub Copilot when external repository content is poisoned,
complementing our focus on deterministic controls at the
configuration-and-process layer upstream of invocation.

NIST AI RMF 1.0 \citep{nist-rmf} and the 2024 Generative AI Profile
\citep{nist-gai} predate widespread tool-using IDE agents; NIST CAISI
launched the \textbf{AI Agent Standards Initiative} (February 2026) to
address identity, authorization, monitoring, and interoperability for
agents \citep{nist-agent}. Berkeley CLTC's \textbf{AI Risk-Management
Standards Profile for GPAI and Foundation Models} (v1.2, 2026) extends
the RMF four-function model with levers relevant to increasingly agentic
systems - human oversight, containment, and accountability for delegated
action \citep{berkeley-agentic}. Rel(AI)Build implements
\textbf{configuration-and-process} controls complementary to these
emerging profiles: it does not duplicate runtime observability
(AgentOps, OpenTelemetry) or output guardrails (Guardrails AI, NeMo
Guardrails).

\subsubsection{2.4 Software supply-chain security and
LLMs}\label{software-supply-chain-security-and-llms}

The intersection of supply chains and LLMs is an emerging threat vector.
Research has shown that AI-recommended packages are frequently
hallucinated \citep{vulcan}, and adversaries actively weaponise these
hallucinations \citep{lanyado}. Ohm et al. \citep{ohm} taxonomise
open-source supply-chain attacks, while Ladisa et al. \citep{ladisa}
extend this taxonomy to cover novel attack vectors including dependency
confusion and CI/CD script injection. Zahan et al. \citep{zahan}
empirically characterise weak links in the npm ecosystem specifically,
finding that a small number of highly-connected packages represent
disproportionate systemic risk - a structural observation that applies
equally to widely-shared agent configuration templates.

Standards bodies are responding: The \textbf{NIST AI RMF}
\citep{nist-rmf} and its \textbf{Generative AI Profile} \citep{nist-gai}
emphasise human oversight, proportional governance, and risk mapping for
generative systems; the \textbf{OWASP Top 10 for LLM Applications}
\citep{owaspllm} highlights excessive agency and supply-chain
vulnerabilities more directly for tool-using agents. NIST's preliminary
\textbf{Cyber AI Profile} (IR 8596, 2025) maps CSF 2.0 cybersecurity
functions to AI-specific risks including agentic threat surfaces
\citep{nist-cyber-ai}. The \textbf{SLSA framework} \citep{slsa} and
\textbf{Sigstore} \citep{sigstore} provide supply-chain integrity
standards for conventional software artifacts. Classic open-source
supply chain incidents - event-stream (2018), Codecov (2021),
ua-parser-js (2021), and xz-utils (CVE-2024-3094) \citep{ohm} - motivate
concrete guardrails (§4.3). Table 4 blocklist entries instantiate Ladisa
et al. \citep{ladisa} attack classes - particularly confused-deputy (T6)
and workspace-trust / CI script injection (T7) - at the
agent-configuration layer. Our contribution extends supply-chain
security upstream to the agent configurations themselves.

\subsubsection{2.5 Prefix caching and prompt
stability}\label{prefix-caching-and-prompt-stability}

Prefix/KV-cache reuse is a high-leverage LLM-serving optimisation
\citep{promptcache}; provider prompt caching (e.g., Anthropic, OpenAI)
rewards stable prompt prefixes. Separately, literature demonstrates that
LLMs are highly sensitive to spurious formatting variations
\citep{sclar, mishra, lu}, where cosmetic formatting differences change
BPE token sequences and push inputs off-distribution. Our architecture
structures content to prioritize static elements for prefix caching
(§4.6) and formalizes formatting drift measurement against a baseline
(§4.4).

\subsubsection{2.6 Capability-based security and least-privilege
enforcement}\label{capability-based-security-and-least-privilege-enforcement}

Least-privilege and capability-based security - from seccomp profiles
\citep{seccomp} to object-capability systems \citep{miller}
\citep{saltzer, dennisvanhorn} - enforce fine-grained allowlists at the
system boundary. Our permission tier model (§4.2) applies this lineage
to agent definitions as compile-time errors on the canonical definition,
not as per-API-call runtime exceptions.

\subsubsection{2.7 Configuration management and infrastructure as
code}\label{configuration-management-and-infrastructure-as-code}

Infrastructure-as-code \citep{humble, morris} and
configuration-management lineage \citep{puppet} treat executable
configuration as versionable, testable software; Rahman et al.
\citep{rahman} show security smells persist when change-control
discipline is absent. Our architecture extends this principle to agent
definitions via a canonical registry, content hashing, and a per-target
policy compiler.

\subsubsection{2.8 Stage-gate models and formal workflow
verification}\label{stage-gate-models-and-formal-workflow-verification}

Cooper's stage-gate model \citep{cooper} and formal workflow
verification - from Petri nets \citep{murata} and process algebras
\citep{hoare} \citep{aalst} - motivate mandatory review checkpoints. Our
phase-gated lifecycle (§4.5) instantiates these as a deterministic state
machine whose gates block progression until invariants are satisfied.

\subsubsection{2.9 Tamper-evident logging and
provenance}\label{tamper-evident-logging-and-provenance}

Hash-chained tamper-evident logging \citep{haberstornetta} - a
construction familiar from append-only ledger designs \citep{nakamoto} -
deployed at scale in Certificate Transparency \citep{ct} - enables
retrospective audit without trusting the log server. Our append-only
audit log (§4.1) applies the same construction locally.

\subsubsection{2.10 Reference currency}\label{reference-currency}

Foundational citations (least-privilege, stage-gate, tamper-evident
logging) are intentional anchors for established paradigms. Fast-moving
LLM and agent-harness references are bounded to our corpus snapshot date
(June 2026); Table 2 notes that harness configuration formats evolve
rapidly.

\begin{center}\rule{0.5\linewidth}{0.5pt}\end{center}

\subsection{3. System Architecture}\label{system-architecture}

The proposed control plane architecture operates in three distinct
planes. Our reference implementation, Rel(AI)Build, instantiates these
planes in Node.js.

\begin{itemize}
\tightlist
\item
  \textbf{Authoring \& Distribution Plane} - A canonical registry of
  agents, skills, knowledge shards, profiles, and workflows, secured
  with content hashes. A CLI and an MCP server expose install, search,
  compose, and governance operations.
\item
  \textbf{Compilation Plane} - A transformer that normalises a canonical
  Markdown+YAML definition and emits IDE-native files for diverse
  targets, explicitly ordering static content first to optimize for
  prompt-cache efficiency.
\item
  \textbf{Runtime-Governance Plane} - Deterministic helpers (CLI scripts
  and/or MCP tools) that the in-IDE agent calls to advance phase state,
  record delegations, and update the requirement→file→test trace. This
  plane enforces permissions and command/path blocklists at install-time
  and pre-tool-use.
\end{itemize}

\begin{center}\includegraphics[width=\linewidth,height=0.6\textheight,keepaspectratio]{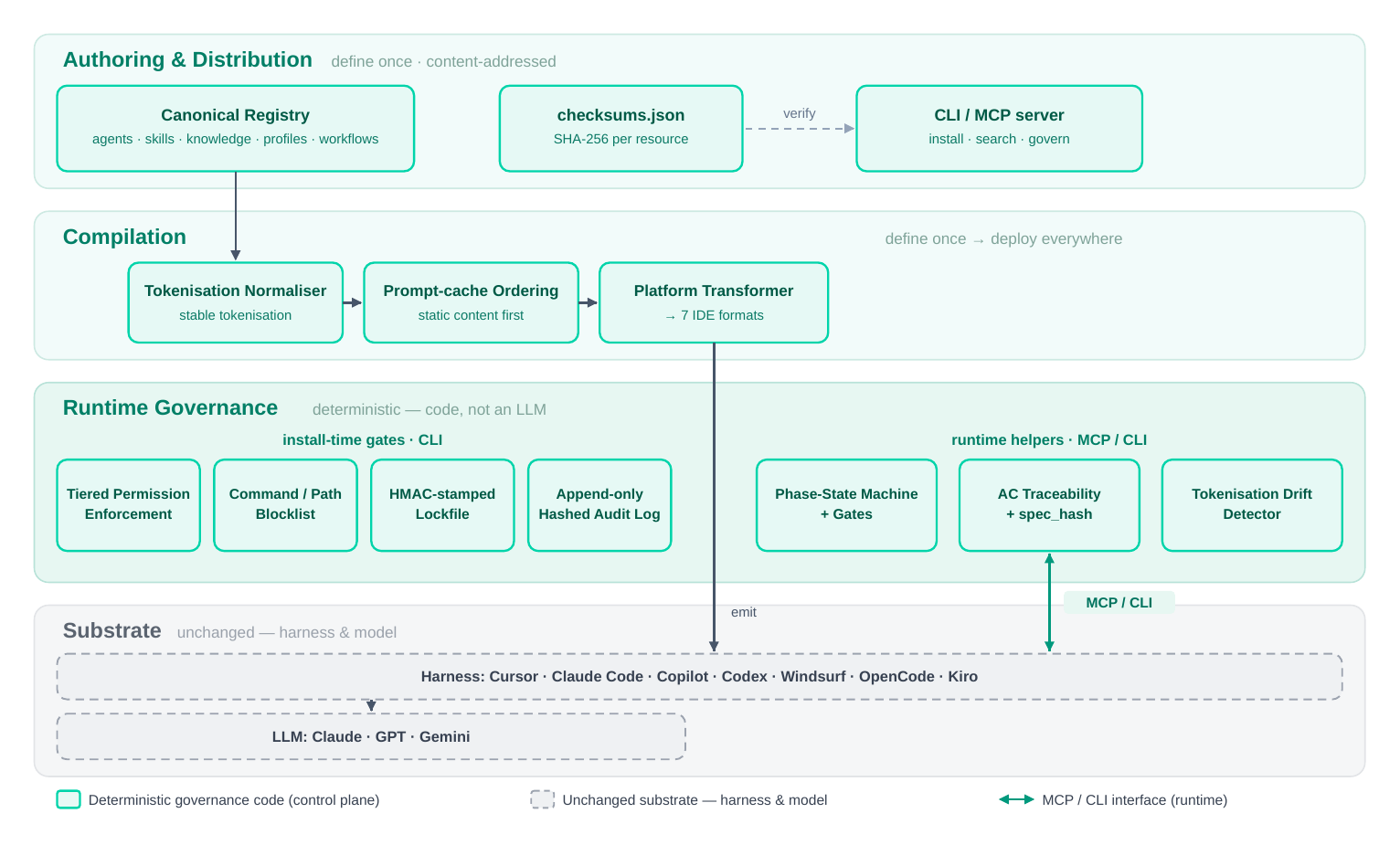}\end{center}

\emph{Figure 2. The deterministic control plane architecture. Four
horizontal layers - Authoring \& Distribution, Compilation, Runtime
Governance (teal), and the unchanged harness/model substrate (grey) -
sit above one another. Every governance component is deterministic code,
not an LLM prompt; the harness and model are unmodified.}

\subsubsection{3.1 The Five Resource
Pillars}\label{the-five-resource-pillars}

The architecture relies on five distinct resource types, separating
concerns that bare configuration files typically conflate:

\begin{itemize}
\tightlist
\item
  \textbf{Agents} - \emph{who}: Role-scoped actors with a system prompt,
  a declared tool allowlist, a permission tier, glob scoping, and a
  model-profile reference.
\item
  \textbf{Skills} - \emph{how}: Reusable, parameterised task recipes an
  agent invokes for a sub-task.
\item
  \textbf{Knowledge shards} - \emph{what to know}: Pre-distilled
  patterns, anti-patterns, and conventions, loaded just-in-time via an
  index rather than bulk-injected.
\item
  \textbf{Profiles} - \emph{standing context}: Always-apply,
  stack-specific rules.
\item
  \textbf{Workflows} - \emph{when}: Multi-step task templates executed
  by an orchestrator.
\end{itemize}

\subsubsection{3.2 Reference
Implementation}\label{reference-implementation}

The reference implementation, Rel(AI)Build, is written in Node.js. The
transformation, permission enforcement, auditing, drift calculation, and
state machine transition logic exist as isolated, independently testable
modules shared between the \texttt{relaibuild} CLI and an MCP server.
This module boundary is deliberate: the CLI is the authoritative
governance path (it can run in any CI environment), while the MCP server
exposes the same operations in-IDE without requiring terminal approval.
A continuous integration validator runs structural checks (tier
membership, tool validity, supply-chain policy coverage) on every
registry change. The isolation of each mechanism into a named module
means that the SHA-256 integrity gate, the HMAC stamp, the state
machine, and the Jaccard calculator are each independently auditable and
replaceable.

\subsubsection{3.3 Runtime Artifacts}\label{runtime-artifacts}

The control plane persists governance state in deterministic files under
\texttt{docs/\{ticket\}/} and at the project root:

{\def\LTcaptype{none} 
\begin{longtable}[]{@{}
  >{\raggedright\arraybackslash}p{(\linewidth - 4\tabcolsep) * \real{0.3333}}
  >{\raggedright\arraybackslash}p{(\linewidth - 4\tabcolsep) * \real{0.3333}}
  >{\raggedright\arraybackslash}p{(\linewidth - 4\tabcolsep) * \real{0.3333}}@{}}
\toprule\noalign{}
\begin{minipage}[b]{\linewidth}\raggedright
Artifact
\end{minipage} & \begin{minipage}[b]{\linewidth}\raggedright
Purpose
\end{minipage} & \begin{minipage}[b]{\linewidth}\raggedright
Written by
\end{minipage} \\
\midrule\noalign{}
\endhead
\bottomrule\noalign{}
\endlastfoot
\texttt{phase-state.json} & Phase lifecycle, delegations, security-scan
receipts, HITL pauses & \texttt{workflow-state} CLI/MCP \\
\texttt{trace.json} & AC → files → tests linkage, \texttt{spec\_hash} &
\texttt{trace-update} CLI/MCP \\
\texttt{spec.md} & Content-addressed requirements (hashed at approval) &
spec-author / human \\
\texttt{handoff.md} & Delegation receipts between agents &
\texttt{write-handoff} CLI/MCP \\
Lockfile (per-project) & Installed resources + HMAC stamp &
\texttt{relaibuild\ install} \\
Audit log (JSONL) & Hash-chained mutation history & CLI/MCP mutations \\
\end{longtable}
}

Subsequent sections reference these artifacts by name; §4.5 and Appendix
C detail phase-state transitions.

\begin{center}\rule{0.5\linewidth}{0.5pt}\end{center}

\subsection{4. Governance Mechanisms}\label{governance-mechanisms}

This section outlines the deterministic mechanisms. We describe the
\emph{what} and \emph{why}, providing pseudocode to clarify the
structural invariants.

\subsubsection{4.1 Agent definitions as a managed supply
chain}\label{agent-definitions-as-a-managed-supply-chain}

\textbf{Content addressing.} Every registry resource possesses a SHA-256
hash. The installation process recomputes the hash of fetched content
and aborts on mismatch - mimicking the integrity model of a package
lockfile, applied to prompts. This mitigates \emph{context poisoning}
(T3, §5).

\textbf{Tamper-evident lockfile.} A per-project lockfile records
installed resources, targets, and origins. On write, it is stamped with
an HMAC-SHA256 over the serialised content. By default, using a
machine-local key derived from the lockfile path, this provides
\textbf{detection of accidental or automated corruption} (merge
conflicts, accidental edits) - not cryptographic authentication against
an adversarial process running as the same user, who can read the key
and forge a valid stamp. Deployments requiring strong authentication can
substitute an organisational signing key (Sigstore-grade) without
altering the mechanism; that upgrade path is not yet implemented in the
reference build.

\textbf{Append-only, hash-chained audit log.} CLI mutations append a
JSONL record \texttt{\{timestamp,\ action,\ user,\ cwd,\ details\}} to a
local audit log, with each line carrying a SHA-256 hash of its own
content. A verifier recomputes every line's hash to report tampering,
yielding a provenance trail.

\begin{Shaded}
\begin{Highlighting}[]
\NormalTok{verify\_audit(log):}
\NormalTok{  for each line i:}
\NormalTok{     \{\_hash, rest\} = parse(line)}
\NormalTok{     if sha256(serialize(rest))[:16] != \_hash: report TAMPERED(i)}
\end{Highlighting}
\end{Shaded}

\begin{center}\includegraphics[width=\linewidth,height=0.85\textheight,keepaspectratio]{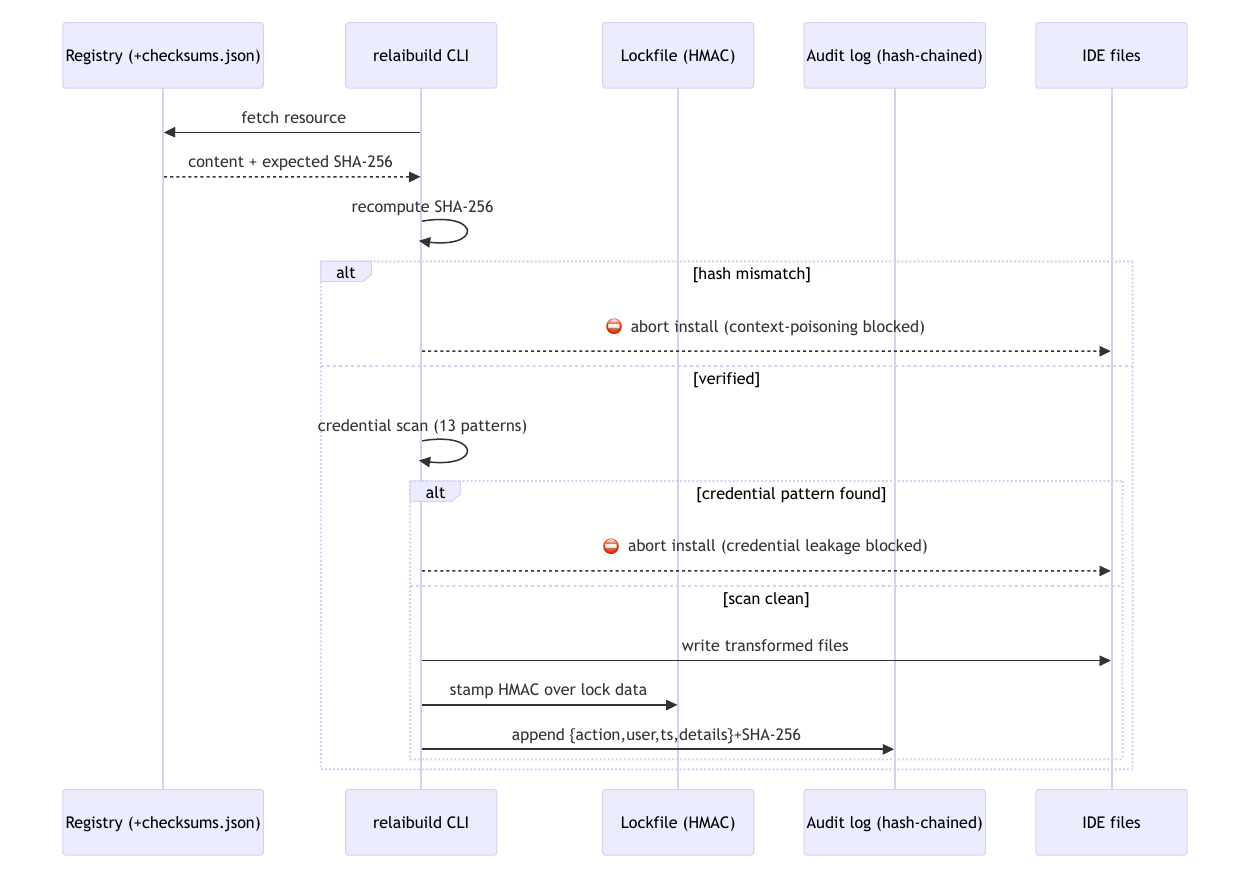}\end{center}

\emph{Figure 3. The agent-definition install pipeline. Three independent
integrity stores (checksum, lockfile, audit log) provide content
integrity, tamper-evidence, and provenance respectively. Two sequential
abort gates - hash mismatch and credential-pattern detection - each
independently block installation, providing defense-in-depth against
context poisoning and credential leakage.}

\subsubsection{4.2 Deterministic permission
model}\label{deterministic-permission-model}

Every agent is assigned a single tier that defines a strict
\emph{allowlist} of tools (Table 3). While the concept of RBAC is
standard, our architecture enforces this deterministically as
\textbf{errors at install/transform time} (fail-closed), ensuring a
misconfigured agent never reaches the IDE execution context.

\textbf{Table 3. Permission tiers (least → most privileged).}

{\def\LTcaptype{none} 
\begin{longtable}[]{@{}
  >{\raggedright\arraybackslash}p{(\linewidth - 8\tabcolsep) * \real{0.2000}}
  >{\raggedright\arraybackslash}p{(\linewidth - 8\tabcolsep) * \real{0.2000}}
  >{\raggedright\arraybackslash}p{(\linewidth - 8\tabcolsep) * \real{0.2000}}
  >{\raggedright\arraybackslash}p{(\linewidth - 8\tabcolsep) * \real{0.2000}}
  >{\raggedright\arraybackslash}p{(\linewidth - 8\tabcolsep) * \real{0.2000}}@{}}
\toprule\noalign{}
\begin{minipage}[b]{\linewidth}\raggedright
Tier
\end{minipage} & \begin{minipage}[b]{\linewidth}\raggedright
Intent
\end{minipage} & \begin{minipage}[b]{\linewidth}\raggedright
Tools (summary)
\end{minipage} & \begin{minipage}[b]{\linewidth}\raggedright
MCP
\end{minipage} & \begin{minipage}[b]{\linewidth}\raggedright
Shell
\end{minipage} \\
\midrule\noalign{}
\endhead
\bottomrule\noalign{}
\endlastfoot
readonly & exploration, review & read/glob/grep/search & allowlisted
only & no \\
scribe & scoped writers (scanner, docs) & read + write/edit
\textbf{within declared paths} & no & no \\
operations & git/CI/deploy & read/write/edit/bash/git & declared only &
yes \\
specialist & coding/analysis & read/write/edit/bash/glob/grep & declared
only & yes \\
orchestrator & top-level coordination & unrestricted & yes & yes \\
\end{longtable}
}

\textbf{Scoped writes with traversal defence.} The enforcement logic
normalises the path (collapsing \texttt{.}/\texttt{..}), checks the
prefix, and explicitly rejects any residual \texttt{..} segments to
prevent directory traversal.

\begin{Shaded}
\begin{Highlighting}[]
\NormalTok{validateWritePath(agent, path):}
\NormalTok{  if tier(agent) != scribe: return}
\NormalTok{  p = normalize(strip\_root(path))            \# collapse . and ..}
\NormalTok{  if ".." in segments(p): throw TRAVERSAL}
\NormalTok{  if not any(p startswith allowed for agent): throw OUT\_OF\_SCOPE}
\end{Highlighting}
\end{Shaded}

\textbf{Agent-level hard constraints.} Beyond tier allowlists, canonical
agent definitions embed \emph{MUST NOT} rules in the system prompt that
hold regardless of task input. The \texttt{git-ops} operations agent -
which has shell access - explicitly forbids \texttt{git\ config}, writes
to \texttt{.git/config} or \texttt{.git/hooks/}, and
\texttt{core.hooksPath} manipulation, and requires branch commands to
run as separate steps rather than \texttt{\&\&}-chained one-liners.
These constraints address workspace-trust attacks (T7, §5) in which
poisoned instructions in an untrusted repository attempt to redirect git
hooks for remote code execution.

\subsubsection{4.3 Attack-derived command and path
blocklists}\label{attack-derived-command-and-path-blocklists}

The architecture utilizes regular expressions to block shell commands
and write targets associated with documented software supply-chain
attacks \citep{ohm}. These blocks apply to \textbf{all} tiers, including
orchestrators. Each pattern traces to a specific incident profile:

\textbf{Table 4. Blocklist patterns ↔ originating incident.}

{\def\LTcaptype{none} 
\begin{longtable}[]{@{}
  >{\raggedright\arraybackslash}p{(\linewidth - 4\tabcolsep) * \real{0.3333}}
  >{\raggedright\arraybackslash}p{(\linewidth - 4\tabcolsep) * \real{0.3333}}
  >{\raggedright\arraybackslash}p{(\linewidth - 4\tabcolsep) * \real{0.3333}}@{}}
\toprule\noalign{}
\begin{minipage}[b]{\linewidth}\raggedright
Pattern (id)
\end{minipage} & \begin{minipage}[b]{\linewidth}\raggedright
Technique blocked
\end{minipage} & \begin{minipage}[b]{\linewidth}\raggedright
Incident Class
\end{minipage} \\
\midrule\noalign{}
\endhead
\bottomrule\noalign{}
\endlastfoot
\texttt{npx-autoconfirm} / \texttt{npm-exec-autoconfirm} &
\texttt{npx\ -y} auto-confirm payload delivery & CI/CD script hijack \\
\texttt{git-sha-fetch}, \texttt{pip-git-sha} & install from arbitrary
commit SHA & Dependency pivot \\
\texttt{pipe-to-sh} & \texttt{curl\ …\ \textbar{}\ bash} payload
delivery & Codecov (2021) \\
\texttt{detached-spawn} & \texttt{spawn(…,\ \{detached:true\})}
self-daemonisation & ua-parser-js (2021) \\
persistence paths & writes to \texttt{cron}, \texttt{systemd}, XDG
\texttt{autostart} & xz-utils (CVE-2024-3094) \\
\texttt{git-config-global} / \texttt{git-config-system} &
\texttt{git\ config\ -\/-global} / \texttt{-\/-system} writes &
Workspace trust / hook injection \\
\texttt{git-hookspath} & \texttt{core.hooksPath} manipulation (redirects
git hooks) & Workspace trust / hook injection \\
\texttt{git-config-file-write} & direct \texttt{.git/config} access &
Workspace trust / hook injection \\
\end{longtable}
}

On \texttt{relaibuild\ install}, the reference implementation copies
\texttt{security-scan.js} into the destination project's IDE-specific
scripts directory (e.g.~\texttt{.cursor/scripts/},
\texttt{.claude/scripts/}) and wires it into \textbf{runtime hooks} that
screen commands \emph{before} execution:

\begin{itemize}
\tightlist
\item
  \textbf{Claude Code:} \texttt{PreToolUse} hooks in
  \texttt{.claude/settings.json} - \texttt{pre-write} (file content) and
  \texttt{pre-exec} (every \texttt{Bash} call).
\item
  \textbf{Cursor:} a \texttt{beforeShellExecution} entry in
  \texttt{.cursor/hooks.json} (\texttt{failClosed:\ true}) invoking
  \texttt{.cursor/hooks/security-pre-exec-scan.sh}, which calls
  \texttt{security-scan.js\ -\/-mode\ pre-exec}.
\end{itemize}

A \textbf{second layer} runs after coding-agent delegations:
orchestrators invoke \texttt{scan-diff} (Step N.2b), checking only
\emph{added} lines in the git diff. The result is persisted to
\texttt{phase-state.json} (§3.3); the phase state machine refuses
\texttt{end-phase} until \texttt{securityScan.status} is \texttt{clean}
when delegations occurred (§4.5). Phases without file-writing
delegations are exempt.

\subsubsection{4.4 Tokenisation normalisation and drift
detection}\label{tokenisation-normalisation-and-drift-detection}

\textbf{Normalisation.} Before compilation, the Markdown body (excluding
code fences or frontmatter) is canonicalised: CRLF→LF,
trailing-whitespace stripping, leading tabs→spaces, and blank-line
collapse. This stabilises the BPE token sequence across platforms to
mitigate model sensitivity to formatting \citep{sclar}.

\textbf{Metric choice: Jaccard similarity.} We fingerprint each agent
body as a set of normalised word tokens and measure drift against a
stored baseline using Jaccard similarity \citep{broder}. The choice of
Jaccard over alternatives warrants justification. Edit distance
(Levenshtein) and cosine-over-TF-IDF are character- or
frequency-weighted measures ill-suited to governance: a single injected
flag such as \texttt{-\/-no-verify} costs exactly one edit step in
Levenshtein yet represents a semantically critical change, whereas
Jaccard set arithmetic treats the \emph{presence or absence} of any
token as a structural signal regardless of position or frequency. This
property - detecting token presence/absence independent of position -
matches the adversarial model of a targeted prompt-injection attack, in
which an attacker typically \emph{adds} a small number of semantically
critical tokens to an otherwise legitimate prompt rather than
reorganising the entire text. Levenshtein's positional and frequency
blindness would systematically under-count such attacks; Jaccard's set
model does not. MinHash approximation of Jaccard is appropriate for
near-duplicate detection at corpus scale; at per-file granularity, exact
Jaccard is O(n) and imposes no accuracy penalty.

The gap between word-level Jaccard and BPE-token-sequence divergence is
intentional. BPE tokenisation is provider- and version-specific;
word-level tokenisation is deterministic, portable, and auditable.
Jaccard operates on sets, making it insensitive to argument order and
whitespace layout - exactly the cosmetic variations that §4.4
normalisation already handles. The metric therefore measures
\emph{lexical content change} independently of BPE implementation
details, and the normalisation step preceding compilation ensures that
formatting noise does not inflate the measured drift.

\[J(A,B)=\frac{|A\cap B|}{|A\cup B|},\qquad \text{risk}=\begin{cases}\text{LOW}& J\ge 0.80\\ \text{MEDIUM}& 0.60\le J<0.80\\ \text{HIGH}& J<0.60\end{cases}\]

\textbf{Threshold defaults.} The boundaries (0.60, 0.80) are
\emph{operating defaults pending calibration (§9.1)}, not empirically
validated against behavioural-change metrics. They were chosen for
operational triage: at J = 0.80, roughly one word in five differs from
the baseline - the volume of a targeted single-instruction edit; at J =
0.60, two words in five differ - approaching a structural rewrite. The
predicted correlation between HIGH drift and measurable behavioural
change is deferred to §9.1. The same Jaccard metric family (default
threshold 0.80) is used in the corpus near-clone analysis of §7;
exact-duplicate propagation (\textbf{10.1\%} fork-adjusted;
\textbf{18.75\%} raw) uses SHA-256 and is threshold-independent.

\textbf{Table 5. Near-clone detection at default threshold (config-file
corpus, Jaccard token sets).}

{\def\LTcaptype{none} 
\begin{longtable}[]{@{}
  >{\raggedright\arraybackslash}p{(\linewidth - 4\tabcolsep) * \real{0.3000}}
  >{\raggedleft\arraybackslash}p{(\linewidth - 4\tabcolsep) * \real{0.4000}}
  >{\raggedright\arraybackslash}p{(\linewidth - 4\tabcolsep) * \real{0.3000}}@{}}
\toprule\noalign{}
\begin{minipage}[b]{\linewidth}\raggedright
Threshold pair (LOW/MED boundary)
\end{minipage} & \begin{minipage}[b]{\linewidth}\raggedleft
Near-duplicate pairs
\end{minipage} & \begin{minipage}[b]{\linewidth}\raggedright
Notes
\end{minipage} \\
\midrule\noalign{}
\endhead
\bottomrule\noalign{}
\endlastfoot
0.60 / 0.80 (default) & 32,677 & Template pairs: 52,833; cross-org rate:
75.51\% (§7) \\
\end{longtable}
}

Sensitivity at other threshold pairs is deferred to corpus completion
(§9.3). Headline exact-duplicate claims (\textbf{10.1\%} fork-adjusted
primary; \textbf{18.75\%} raw) use SHA-256 and are
threshold-independent. Threshold sensitivity against agent behavioural
change is not empirically validated; see §8 and §9.1.

\begin{center}\includegraphics[width=\linewidth,height=0.6\textheight,keepaspectratio]{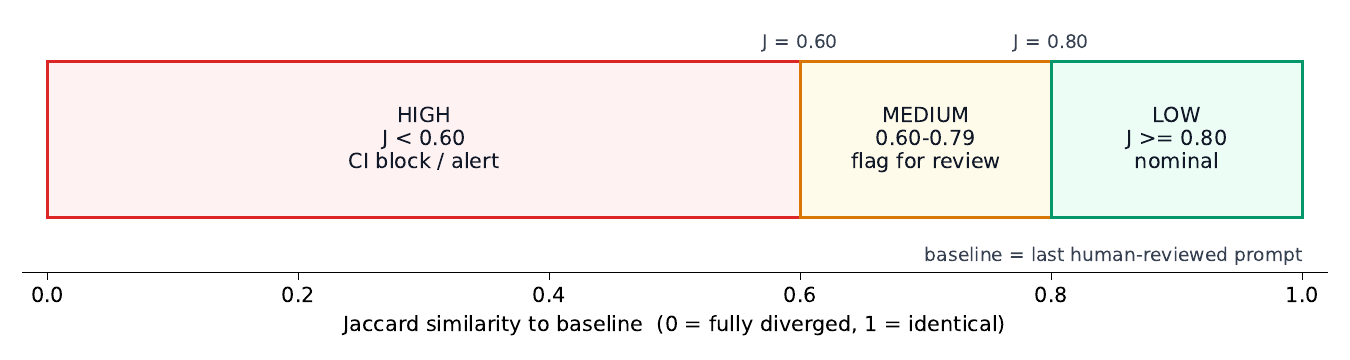}\end{center}

\emph{Figure 4. Conceptual illustration of tokenisation drift risk
zones. Thresholds shown (J \textless{} 0.60 = HIGH, 0.60--0.79 = MEDIUM,
≥ 0.80 = LOW) are operating defaults pending calibration (§9.1). The §7
corpus study uses J ≥ 0.80 for near-clone detection. Baseline is the
last human-reviewed version. A HIGH result triggers a CI-failing signal
requiring human review.}

\textbf{Table 5b. Illustrative Jaccard sensitivity to targeted token
injection (no behavioural validation; thresholds are operating defaults
per §9.1).}

Model: baseline prompt of \emph{n} unique word tokens; attacker injects
\emph{k} new tokens not present in the baseline; (J = n/(n+k)).

{\def\LTcaptype{none} 
\begin{longtable}[]{@{}lrrrr@{}}
\toprule\noalign{}
Baseline length \emph{n} \textbackslash{} injected tokens \emph{k} & 1 &
3 & 5 & 10 \\
\midrule\noalign{}
\endhead
\bottomrule\noalign{}
\endlastfoot
50 & 0.980 & 0.943 & 0.909 & 0.833 \\
100 & 0.990 & 0.971 & 0.952 & 0.909 \\
200 & 0.995 & 0.985 & 0.976 & 0.952 \\
\end{longtable}
}

A single-token injection into a 100-token baseline yields (J
\approx 0.99) - above the LOW threshold (≥0.80) and therefore not
flagged as drift. The HIGH-risk zone (\textless0.60) requires
substantially larger lexical edits (roughly ≥40\% new tokens relative to
baseline length under this additive model), motivating why drift
detection targets multi-token or structural prompt changes rather than
isolated flag tokens alone.

\subsubsection{4.5 Deterministic phase-gated lifecycle with hard
invariants}\label{deterministic-phase-gated-lifecycle-with-hard-invariants}

Feature work is structured into a bounded, sequential pipeline with
mandatory human-in-the-loop (HITL) gates and a hard \textbf{3-iteration
cap} on auto-fix loops \citep{ddi, nist-gai, berkeley-agentic}. State
persists to \texttt{phase-state.json} (§3.3), enabling exact
resumption.\footnote{Phase numbering (0, 1, 3, 4, 5, 7, 8) is
  non-contiguous by design: phases 2 and 6 are reserved for future
  specialist sub-workflows without renumbering deployed systems.}

Crucially, a \textbf{deterministic state machine enforces invariants and
blocks on violation}:

\begin{itemize}
\tightlist
\item
  \textbf{Ordering invariant.} Starting phase \emph{n} (\textgreater1)
  fails unless phase \emph{n−1} recorded an end timestamp.
\item
  \textbf{End-without-start invariant.} Ending a phase that never
  started fails.
\item
  \textbf{Delegation receipts.} Ending a phase can require named
  receipts (\texttt{DELEGATION\ GATE\ FAILURE}), preventing an
  orchestrator from silently bypassing specialist delegation.
\item
  \textbf{Mandatory review gate.} For critical tiers, phase 0 cannot
  close unless a \texttt{spec-reviewer} delegation is recorded.
\item
  \textbf{Post-delegation security scan gate.} When file-writing
  delegations were recorded for phase \emph{n}, ending the phase fails
  unless \texttt{phase\_n.securityScan.status\ ==\ clean} (or no
  delegations occurred - pure HITL/planning phases are exempt). This
  enforces orchestrator Step N.2b at the CLI/MCP layer, not only in
  protocol text.
\end{itemize}

\begin{Shaded}
\begin{Highlighting}[]
\NormalTok{end\_phase(n):}
\NormalTok{  if not metrics[n].start\_ts: fail END\_WITHOUT\_START}
\NormalTok{  if required\_delegations not ⊆ recorded[n]: fail DELEGATION\_GATE}
\NormalTok{  if n==0 and tier∈\{M,L\} and \textquotesingle{}spec{-}reviewer\textquotesingle{} ∉ recorded[0]: fail REVIEW\_GATE}
\NormalTok{  if delegations[n] ≠ ∅ and securityScan[n].status ≠ clean: fail SECURITY\_SCAN\_GATE}
\NormalTok{  commit(end\_ts, active = total − hitl\_pause)}
\end{Highlighting}
\end{Shaded}

Because these checks are hard-coded in the control plane, they cannot be
bypassed via persuasive LLM prompting.

\textbf{Rejection and rollback mechanics.} The forward invariants above
are only half the story; HITL rejection paths are equally deterministic.
When a human reviewer rejects at a HITL gate (e.g., Phase 5 Review), the
state machine records the rejection in the append-only audit log as a
\texttt{HITL\_REJECTED} event with a timestamp and reviewer identifier,
and leaves the preceding phase's \texttt{end\_ts} intact but marks it as
\emph{pending re-entry}. The rejected work item returns to the start of
the phase that was just reviewed - the state machine does not revert
further back (e.g., a failed Phase 5 Review does not revert to Phase 3
Implementation). The 3-iteration cap on auto-fix loops is scoped to a
single phase entry: a rollback from Phase 5 back into Phase 4 resets the
Phase 4 counter to zero, because the reviewer's rejection constitutes a
new mandate. All backward transitions - rejection events and re-entries
- are appended to the audit log using the same append-only, hash-chained
format as forward transitions, ensuring a complete and tamper-evident
history of the full lifecycle including rework cycles.

\begin{center}\includegraphics[width=\linewidth,height=0.85\textheight,keepaspectratio]{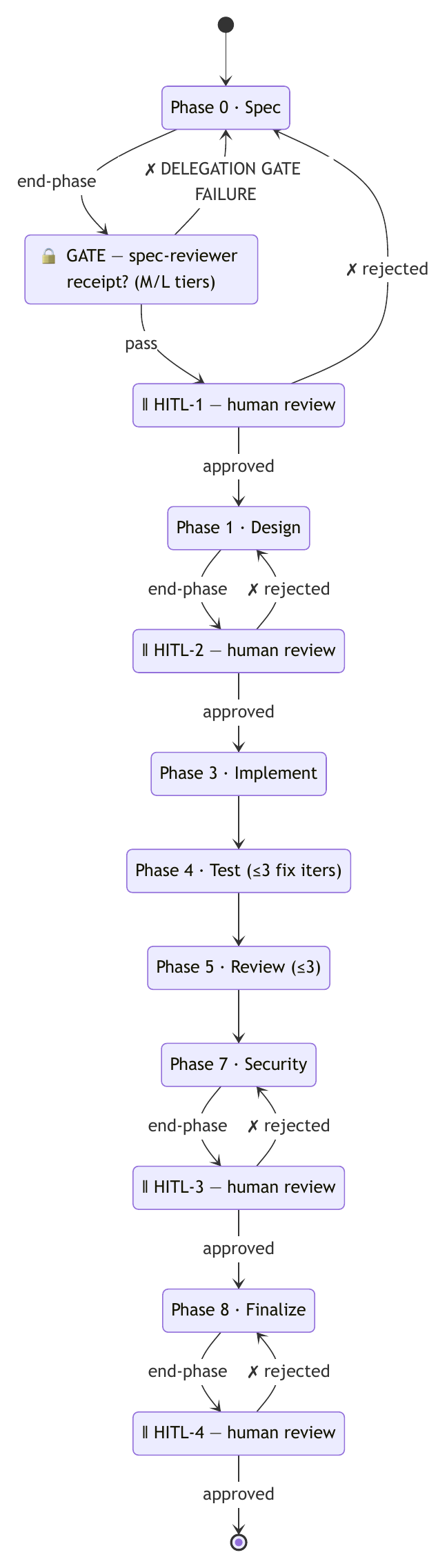}\end{center}

\emph{Figure 5. The deterministic phase-gated lifecycle. Four HITL gates
(pause bars) enforce mandatory human review; gate-failure back-edges
(dashed, red) refuse advancement until invariants are satisfied. Three
hard invariants hold throughout: (1) phase n cannot start until phase
n−1 has ended, (2) ending a phase that never started is blocked, and (3)
auto-fix loops are capped at three iterations before escalating to a
human. Phase numbering is non-contiguous \footnote{Phase numbering (0,
  1, 3, 4, 5, 7, 8) is non-contiguous by design: phases 2 and 6 are
  reserved for future specialist sub-workflows without renumbering
  deployed systems.}.}

\paragraph{Lifecycle design rationale}\label{lifecycle-design-rationale}

The eight-phase decomposition is not arbitrary: it maps Cooper's
stage-gate model \citep{cooper} onto software-engineering artifact
boundaries already used in the reference orchestrators
(\texttt{PHASE\_NAMES} in the workflow state machine - Spec, Design,
Implementation, Test, Review, Security, Finalize). Phases \textbf{2}
(Foundation \& Scaffolding) and \textbf{6} (Static Code Analysis) are
\textbf{reserved slots} for specialist sub-workflows that can be
activated without renumbering deployed \texttt{phase-state.json} files.
Four HITL gates correspond to spec approval (after Phase 0), design
approval (after Phase 1), pre-release review (after Phase 7), and final
merge (after Phase 8) - not to arbitrary process overhead (Figure 5 maps
Cooper gates to Rel(AI)Build phases).

We do not claim that eight phases are uniquely optimal - a five-phase or
twelve-phase decomposition could enforce similar invariants. What the
reference implementation requires is \textbf{bounded recursion}
(3-iteration cap), \textbf{mandatory delegation receipts}, and
\textbf{hard-coded gate predicates} (Appendix C summary) - properties
that a thinner process cannot enforce deterministically. The default cap
of three iterations is an \textbf{operating default}, not a claim of
universal optimality: empirical work on LLM code debugging shows
effectiveness decays sharply within 2--3 attempts \citep{ddi}, and
governance standards require human escalation when autonomous action
fails \citep{nist-gai, berkeley-agentic}. Deployments may override the
default via \texttt{PACK\_MAX\_ITERATIONS} (range 1--5). Phase-gate
\textbf{overhead is unmeasured} (§9.2); necessity is argued from process
invariants, not from productivity data.

\textbf{Requirement→file→test traceability.} A companion
\texttt{trace.json} (§3.3) records, per acceptance criterion (AC), the
implementing files, performed tasks, verifying tests, and status. The
specification is content-hashed at approval. Companion checks verify
that every AC maps to code and tests, designing the system to flag any
files changed outside of AC scopes as detectable scope creep.

\paragraph{Trust boundary: cooperative trace
linkage}\label{trust-boundary-cooperative-trace-linkage}

File-to-AC linkage is established through \emph{explicit agent
declaration}: when a coding agent implements or touches a file for a
given AC, it calls the \texttt{trace-update} CLI command (or equivalent
MCP tool) with the AC identifier and the file path, which appends the
mapping to the trace record for that phase. This is not inferred from
git history; it requires the agent to invoke the trace tool during
implementation. The spec-drift check then computes the set of all files
changed in the working tree (via git diff against the phase-start
commit) and flags any file that does not appear in any AC's registered
file list.

This creates a \textbf{determinism boundary} central to the
architecture's honesty: trace linkage depends on the agent's cooperative
invocation of the trace tool. A misconfigured, prompt-injected, or
non-compliant agent that edits files without calling
\texttt{trace-update} will have its changes surface as scope-creep
warnings at the next drift check - \textbf{detectable but not
preventable} before the edit lands. In a threat model where the same LLM
can be steered by injected instructions, relying on that agent to
faithfully call \texttt{trace-update} is an honor-system component
inside an otherwise deterministic control plane. The control plane's
response is auditability and detection, not pre-execution file-scope
enforcement. A DRIFT-detection vignette (Appendix D) shows the dominant
trace gap is \emph{implementation without test trace} - code
implementing an AC exists but no verifying test is registered (AC-018) -
precisely the failure mode this linkage is designed to surface.

\begin{center}\includegraphics[width=\linewidth,height=0.6\textheight,keepaspectratio]{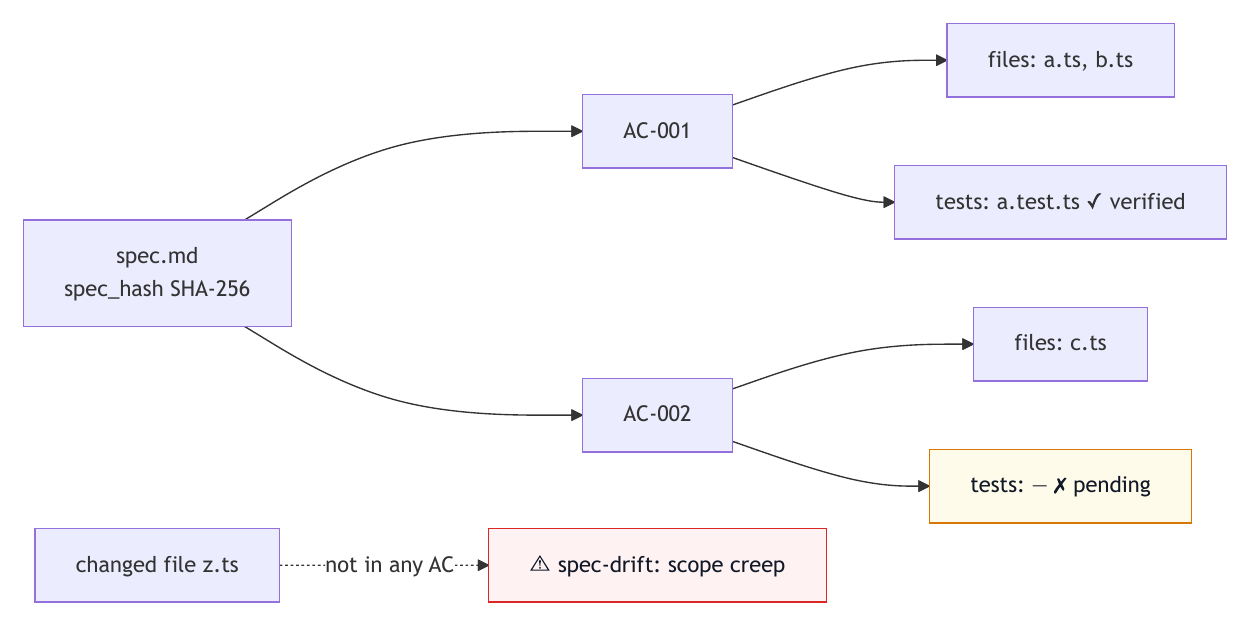}\end{center}

\emph{Figure 6. Requirement→file→test traceability. Each acceptance
criterion (AC) in the content-addressed spec maps to implementing files
and verifying tests; a verified AC is green, an unverified AC is amber.
A file changed outside any AC scope surfaces as a spec-drift warning
(red), making scope creep detectable automatically. File-to-AC linkage
requires cooperative agent invocation of \texttt{trace-update} (§4.5
trust boundary).}

\subsubsection{4.6 Define once, compile to many
targets}\label{define-once-compile-to-many-targets}

A single canonical definition is compiled - after normalisation and
static-content-first ordering - into specific formats for Cursor, Claude
Code, VS Code, Codex, Windsurf, OpenCode, and Kiro. The research
question is whether compilation \emph{preserves governance properties}
across heterogeneous formats. We argue that because the compiler is a
pure function of the canonical definition \emph{D}, tier checks and
traversal defence run before compilation, and the HMAC stamp references
\emph{D}'s content hash rather than per-target output, every target that
faithfully encodes \emph{D}'s permission declarations inherits the
governance property by construction (conditional property-preservation
argument in Appendix B). Whether each harness's \emph{runtime}
enforcement of the emitted permissions is equivalent is harness-specific
(Table 2) and outside this architecture's compile-time scope.

\textbf{IDE substrate enforcement disparities.} To be explicit about
this gap: the control plane \emph{guarantees that the canonical
definition correctly declares its permissions}, not that the target
harness will enforce them identically at runtime. The targets span a
wide enforcement spectrum. Claude Code is the most explicit: it parses
the comma-separated tool allowlist at session start and rejects
invocations outside the declared set, and runs \texttt{PreToolUse}
security hooks on every write and shell invocation. Cursor enforces tool
access implicitly through glob-scoped rule application and MCP server
configuration, and - on install - deploys a
\texttt{beforeShellExecution} hook that fail-closes on dangerous shell
commands (§4.3), partially closing the runtime gap for the highest-risk
tool class even though Cursor has no compile-time tool allowlist check
on the agent file itself. Several targets - Codex, Windsurf, OpenCode,
and Kiro - have little or no runtime permission enforcement at the
agent-definition level; the compiled output correctly declares the
intent, but the harness does not actively block out-of-scope tool use.
This disparity is a known limitation (§8) and a primary motivation for
the pre-execution governance model: the control plane's checks run
\emph{before} the harness has a chance to not enforce them.

\begin{center}\rule{0.5\linewidth}{0.5pt}\end{center}

\subsection{5. Threat Model}\label{threat-model}

We adopt an attacker model where an adversary can influence (a)
untrusted task inputs (tickets, code comments), (b) the shared
agent-system directory via Pull Requests, (c) dependency suggestions,
and/or (d) \textbf{untrusted workspace content} - e.g.~a developer opens
a cloned third-party repository whose \texttt{AGENTS.md},
\texttt{.cursor/rules}, or source comments contain embedded instructions
aimed at agents with shell access. We map the architectural mitigations
to the OWASP guidelines.

\textbf{Table 6. Threats and deterministic mitigations.}

{\def\LTcaptype{none} 
\begin{longtable}[]{@{}
  >{\raggedright\arraybackslash}p{(\linewidth - 6\tabcolsep) * \real{0.2500}}
  >{\raggedright\arraybackslash}p{(\linewidth - 6\tabcolsep) * \real{0.2500}}
  >{\raggedright\arraybackslash}p{(\linewidth - 6\tabcolsep) * \real{0.2500}}
  >{\raggedright\arraybackslash}p{(\linewidth - 6\tabcolsep) * \real{0.2500}}@{}}
\toprule\noalign{}
\begin{minipage}[b]{\linewidth}\raggedright
Threat
\end{minipage} & \begin{minipage}[b]{\linewidth}\raggedright
OWASP mapping
\end{minipage} & \begin{minipage}[b]{\linewidth}\raggedright
Primary deterministic mitigation (§)
\end{minipage} & \begin{minipage}[b]{\linewidth}\raggedright
Residual Risk
\end{minipage} \\
\midrule\noalign{}
\endhead
\bottomrule\noalign{}
\endlastfoot
T1 Prompt injection & LLM01 & HITL review gates + AC-scope flagging
(§4.5); boundary markers & Medium \\
T2 Privilege escalation & LLM06 & Fail-closed tier allowlists; scoped
writes + traversal defence (§4.2) & Low \\
T3 Context poisoning & LLM03 / Supply Chain & SHA-256 integrity + HMAC
lockfile + audit log (§4.1) & Low--Med \\
T4 Credential leakage & LLM02 / LLM06 & 13-pattern scanner on
reads/writes; OS-keychain retrieval & Low \\
T5 Uncontrolled recursion & LLM06 & 3-iteration cap; bounded delegation
depth; sequential phases (§4.5) & Low \\
T6 Confused deputy & Supply Chain & Attack-derived command/path
blocklist; package-manager wrappers (§4.3) & Low--Med \\
T7 Workspace trust / git hook injection & LLM01 / LLM06 &
\texttt{git-ops} hard constraints (no \texttt{git\ config}, no
\texttt{core.hooksPath}); pre-exec blocklist + IDE runtime hooks (§4.3);
HITL branch confirmation; post-delegation \texttt{scan-diff} gate (§4.5)
& Low--Med \\
\end{longtable}
}

\textbf{Residual Risk Note:} Prompt injection (T1) remains
\emph{Medium}. A highly sophisticated, AC-compliant injection might pass
automated scope checks, relying entirely on the human HITL gate. Context
poisoning (T3) is \emph{Low--Medium}: SHA-256 integrity is strong, but
the default HMAC lockfile detects accidental corruption only - not
adversarial forgery by a same-user process (§4.1, §8). Workspace trust
(T7) remains \emph{Low--Medium}: runtime hooks and agent constraints
block the most direct \texttt{git\ config} / \texttt{core.hooksPath}
path, but instructions in an untrusted workspace can still influence
model behaviour at runtime; developers should treat cloned third-party
repositories as untrusted workspaces. The architecture does not claim to
solve prompt injection; it aims to \emph{bound the blast radius}
deterministically.

\begin{center}\includegraphics[width=\linewidth,height=0.85\textheight,keepaspectratio]{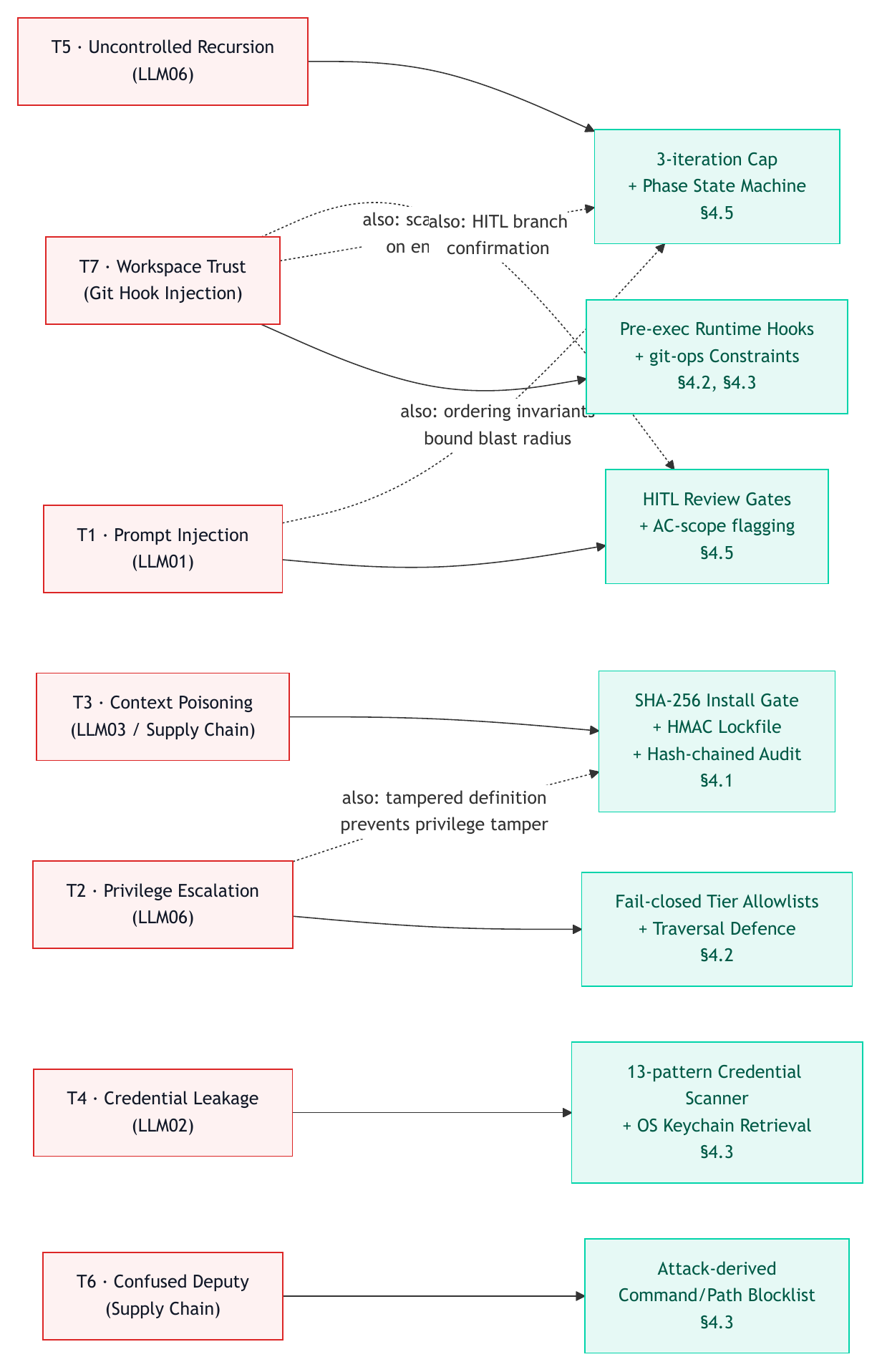}\end{center}

\emph{Figure 7. Threat → deterministic control mapping. Seven identified
threats (red, left) each have a primary deterministic control (teal,
right); T-numbers correspond to Table 6. Cross-links (dashed) show
defense-in-depth: the Phase State Machine (§4.5) also bounds the blast
radius of prompt injection (T1) via ordering invariants and enforces the
post-delegation security scan for T7; SHA-256 supply-chain integrity
(§4.1) also prevents privilege escalation via tampered definitions (T2).
Residual risks for T1 and T7 are noted in Table 6.}

\subsubsection{5.1 Compliance and Supply-Chain
Framing}\label{compliance-and-supply-chain-framing}

The \textbf{NIST AI RMF 1.0} \citep{nist-rmf} organises AI risk
governance across GOVERN, MAP, MEASURE, and MANAGE. Table 7 maps each
function to Rel(AI)Build mechanisms; the 2024 GenAI Profile
\citep{nist-gai} and emerging agentic-AI extensions
\citep{berkeley-agentic, nist-agent} provide adjacent risk vocabulary
but do not prescribe agent-definition supply-chain or phase-traceability
controls - the gap this architecture addresses. Sub-function detail
aligns with AI RMF 1.0 and the GenAI Profile per \citep{companion}.
Organisations mapping controls to ISO/IEC 27001 \citep{iso27001} will
find sub-function detail in \citep{companion}. Executive Order 14028
\citep{eo14028} and the NTIA minimum SBOM elements \citep{ntia} were
developed for conventional software artifacts; to our knowledge, no
prior work has proposed treating agent configurations as SBOM-subject
artifacts. The measured rate of exact-duplicate configuration
propagation (\textbf{10.1\%} fork-adjusted; \textbf{18.75\%} raw; 702
cross-organisation exact clusters, §7) demonstrates that agent configs
already circulate as undeclared shared components - the failure mode
SBOM mandates were designed to surface.

\textbf{Table 7. Top-level NIST AI RMF function alignment.}
\emph{Function names from NIST AI 100-1 (2023) \citep{nist-rmf}.}

{\def\LTcaptype{none} 
\begin{longtable}[]{@{}
  >{\raggedright\arraybackslash}p{(\linewidth - 4\tabcolsep) * \real{0.3333}}
  >{\raggedright\arraybackslash}p{(\linewidth - 4\tabcolsep) * \real{0.3333}}
  >{\raggedright\arraybackslash}p{(\linewidth - 4\tabcolsep) * \real{0.3333}}@{}}
\toprule\noalign{}
\begin{minipage}[b]{\linewidth}\raggedright
AI RMF Function
\end{minipage} & \begin{minipage}[b]{\linewidth}\raggedright
Core Requirement
\end{minipage} & \begin{minipage}[b]{\linewidth}\raggedright
Rel(AI)Build Mechanism
\end{minipage} \\
\midrule\noalign{}
\endhead
\bottomrule\noalign{}
\endlastfoot
GOVERN & Establish policies, accountability structures, and
organisational practices for AI risk & Five-tier permission model
(§4.2); delegation receipts and HITL gates; append-only audit log
(§4.1) \\
MAP & Identify, characterise, and prioritise AI-specific risks in
context & Threat model (Table 6); attack-derived blocklist (§4.3); five
ungoverned risk surfaces (§1.1) \\
MEASURE & Develop and apply metrics to monitor and quantify AI risk
indicators & Jaccard drift detector with operating defaults pending
calibration (§4.4); AC trace coverage (§4.5); hash-chain verifier
(§4.1) \\
MANAGE & Prioritise and implement risk responses; track residual risk &
Phase-gated lifecycle (§4.5); fail-closed permission enforcement (§4.2);
3-iteration escalation cap \\
\end{longtable}
}

\textbf{Table 8. Agent Configuration SBOM: NTIA Minimum Elements →
Rel(AI)Build equivalent.}

{\def\LTcaptype{none} 
\begin{longtable}[]{@{}
  >{\raggedright\arraybackslash}p{(\linewidth - 4\tabcolsep) * \real{0.3333}}
  >{\raggedright\arraybackslash}p{(\linewidth - 4\tabcolsep) * \real{0.3333}}
  >{\raggedright\arraybackslash}p{(\linewidth - 4\tabcolsep) * \real{0.3333}}@{}}
\toprule\noalign{}
\begin{minipage}[b]{\linewidth}\raggedright
NTIA SBOM Element
\end{minipage} & \begin{minipage}[b]{\linewidth}\raggedright
Traditional Software
\end{minipage} & \begin{minipage}[b]{\linewidth}\raggedright
Rel(AI)Build Agent Manifest
\end{minipage} \\
\midrule\noalign{}
\endhead
\bottomrule\noalign{}
\endlastfoot
Supplier name & Package publisher & Registry owner; agent author
metadata in frontmatter \\
Component name & Package / library name & Agent identifier (\texttt{id}
field in canonical frontmatter) \\
Version of the component & Semantic version string & Content-addressed
SHA-256 hash of the canonical definition \\
Dependency relationships & Transitive \texttt{require} / \texttt{import}
graph & \texttt{skills}, \texttt{knowledge}, \texttt{profiles}
references in agent frontmatter \\
Cryptographic hash & SHA-256 of the binary artifact & SHA-256 of the
canonical definition at install time (\texttt{checksums.json}) \\
Unique identifier & PURL / CPE string & Registry-qualified agent ID
concatenated with content hash \\
Timestamp & Build or release timestamp & Append-only audit log entry
timestamp per install or update event \\
\end{longtable}
}

The \texttt{checksums.json} registry, HMAC-stamped lockfile, and
append-only audit log provide integrity and provenance properties
equivalent to CycloneDX/SPDX evidence fields; SPDX/CycloneDX
serialisation is a near-term target. The architecture aligns with
\textbf{SLSA Build Level 2} \citep{slsa}: versioned canonical registry,
scripted deterministic transformer, and version-controlled pipeline.

\subsection{6. Implementation Conformance
Testing}\label{implementation-conformance-testing}

§6 answers \emph{does the mechanism reject bad inputs?} -
enforcement-property tests on the reference registry corpus using
injected violations. This is \textbf{implementation conformance
testing}, not an evaluation of developer outcomes or field
effectiveness. §7 quantifies the public-repository governance gap that
motivates the architecture. A DRIFT-detection vignette on a real API
build (N=1, governed arm only) is provided in Appendix D; it does not
constitute a comparative evaluation.

\subsubsection{6.1 Conformance Corpus and
Methodology}\label{conformance-corpus-and-methodology}

§6 tests mechanism invariants on the reference registry corpus. §7
provides the empirical prevalence measurement.

The conformance corpus consisted of 237 canonical agent definitions
drawn from the reference registry, spanning 14 technology stacks
(Angular, React, Next.js, Vue, React-Native/Expo, Node/Express,
Java/Spring Boot, .NET ASP.NET Core, .NET WPF/MVVM, Python, PHP, SST,
Salesforce, and API-testing) compiled across 7 target IDE formats
(Cursor, Claude Code, VS Code/Copilot, Codex, Windsurf, OpenCode, Kiro).
Simulated policy violations were generated by three methods: (a)
modifying resource content on disk after checksum recording to simulate
registry tampering; (b) adding disallowed tool declarations to
\texttt{scribe} and \texttt{readonly} agent frontmatter to simulate
misconfiguration; and (c) constructing representative shell command
strings and file paths matching the patterns in Table 4 to confirm
blocklist coverage. These tests were executed against the reference
implementation's published CLI and MCP server. The test corpus and
reproduction scripts are included in the artifact appendix.

\subsubsection{6.2 Conformance results}\label{conformance-results}

All 237 canonical definitions compiled cleanly across the 7 target
formats with no schema violations. Each injected violation was caught by
its corresponding gate: all \textbf{10} integrity-tampering cases
aborted at the SHA-256 install gate, all \textbf{15} over-privileged
tool declarations failed compilation under fail-closed tier checks, and
all \textbf{20} synthetic blocklist strings triggered their controls. We
stress what these numbers do and do not mean: they are
\textbf{conformance} checks confirming that hand-written detectors match
inputs constructed to be detected, and that a hash check detects a
changed hash. They carry no information about field effectiveness or
developer outcomes (§8, §9.2); we report them only to establish that
each mechanism enforces its stated invariant. For completeness, the gate
behaviours exercised are: modifying any downloaded resource aborts at
the SHA-256 install gate (editing the lockfile triggers a hash-mismatch
warning on next read); a \texttt{scribe}/\texttt{readonly} agent
declaring a shell tool is a fatal compilation error;
\texttt{validateWritePath} rejects \texttt{../}-masked paths inside an
allowed prefix; and the state machine raises
\texttt{PHASE\ ORDER\ VIOLATION}, \texttt{DELEGATION\ GATE\ FAILURE},
and \texttt{SECURITY\ SCAN\ GATE} on out-of-order, receipt-less, or
unscanned transitions.

\subsection{7. Empirical Prevalence
Study}\label{empirical-prevalence-study}

\begin{quote}
\textbf{Sample adequacy:} ADEQUATE - N=10,008 repositories analysed
across 9,357 distinct organisations meet the target of 10,000;
confidence intervals are narrow enough to support the directional claims
below. Dataset and reproduction scripts are published as a companion
artifact \citep{artifact}.
\end{quote}

To ground the qualitative claims of §1.1, we conducted a preliminary
empirical prevalence study of publicly accessible GitHub repositories.
The study does not evaluate Rel(AI)Build; it measures the governance gap
that the architecture is designed to address.

\subsubsection{7.1 Corpus and methodology}\label{corpus-and-methodology}

Full reproduction parameters are in Appendix A and \citep{artifact}.
Summary:

\textbf{Discovery.} Repositories were assembled via a custom GitHub
corpus analysis pipeline \citep{artifact} using a seed list built from
GitHub Search API queries stratified by star-count quartiles (Q1: 0--10
stars; Q2: 11--100; Q3: 101--1,000; Q4: \textgreater1,000), each
requiring at least one tracked configuration path and
\texttt{pushed:\textgreater{}2024-01-01}. Representative query
templates: \texttt{filename:CLAUDE.md\ stars:0..10},
\texttt{filename:.cursorrules\ stars:11..100},
\texttt{path:.github/workflows\ stars:101..1000},
\texttt{filename:copilot-instructions.md\ stars:\textgreater{}1000}. The
seed list was deduplicated by \texttt{owner/repo} full name. Fork
metadata (\texttt{isFork}, \texttt{parentFullName}) was backfilled for
8,988/10,008 completed repos via the GitHub REST API
(\texttt{enrich-fork-metadata}; 653 fork repos identified).

\textbf{Tracked paths.} \texttt{.github/workflows/*.yml},
\texttt{.github/copilot-instructions.md}, \texttt{.cursor/rules/**},
\texttt{.cursorrules}, \texttt{.claude/**}, and \texttt{CLAUDE.md}.
Files shorter than 100 bytes were excluded as empty stubs.

\textbf{Corpus snapshot.} We analysed 10,008 repositories (373 failed
ingestion; 3.59\% ingestion failure rate), 33,620 files, and 6,145 AI
agent configuration files across 9,357 organisations. Only 8.45\% of
repos (846/10,008) contain at least one AI agent config; 49.0\% contain
CI/CD workflows. Agent-governance claims use \emph{file-level} metrics
as the primary lens (§7.2); repo-level rows in Table 10 are
supplementary.

\textbf{Ingestion failures.} Of 373 failed repositories, per-repo
failure logs were not persisted in the rendered report; failure
characterisation is aggregate only. Failures are attributed to GitHub
API rate limits, repository unavailability (404/archived), and clone
timeouts during enrichment. See Appendix A.

\textbf{Measurements.}

\begin{itemize}
\tightlist
\item
  \emph{Version-control depth:} lifetime commit count per configuration
  file. Files with ≤1 commit are classified as \emph{unmanaged}.
\item
  \emph{Clone propagation:} SHA-256 exact-duplicate groups
  (threshold-independent) with fork-adjusted variants excluding groups
  where fewer than two non-fork repositories share a hash;
  near-duplicate pairs (Jaccard ≥ 0.80 default; Table 5).
\item
  \emph{Credential exposure:} the same 13-pattern regex scanner used at
  install time in §4.3 (metadata only; secret values never stored).
\item
  \emph{Permission scope:} regex extraction of declared tool allowlists
  and GitHub Actions \texttt{permissions} blocks. Parser false positives
  on agent-config formats were manually excluded (6 of 19 raw matches);
  a full Rahman-style validation sample is deferred to corpus completion
  (§9.3).
\end{itemize}

\textbf{Maintenance categories.} \emph{Dormant:} no commit in 18 months
or zero lifetime commits. \emph{Lightly maintained:} 1--5 lifetime
commits and not dormant. \emph{Actively maintained:} ≥6 lifetime commits
and not dormant.

\subsubsection{7.2 Results}\label{results}

\textbf{Table 10. Prevalence study results.} File-level rows are primary
for version-control depth; repo-level rows are supplementary.

{\def\LTcaptype{none} 
\begin{longtable}[]{@{}
  >{\raggedright\arraybackslash}p{(\linewidth - 4\tabcolsep) * \real{0.3333}}
  >{\raggedright\arraybackslash}p{(\linewidth - 4\tabcolsep) * \real{0.3333}}
  >{\raggedright\arraybackslash}p{(\linewidth - 4\tabcolsep) * \real{0.3333}}@{}}
\toprule\noalign{}
\begin{minipage}[b]{\linewidth}\raggedright
Measurement
\end{minipage} & \begin{minipage}[b]{\linewidth}\raggedright
Finding
\end{minipage} & \begin{minipage}[b]{\linewidth}\raggedright
95\% CI / notes
\end{minipage} \\
\midrule\noalign{}
\endhead
\bottomrule\noalign{}
\endlastfoot
{[}file-level{]} AI agent config files with exactly 1 commit & 58.45\%
(3,592/6,145 files) & 57.22\% -- 59.68\% \\
{[}file-level{]} AI agent config files with ≤2 commits & 74.26\%
(4,563/6,145 files) & 73.15\% -- 75.33\% \\
{[}file-level{]} Median commit count per AI agent config file & 1 & IQR:
1--3 (\emph{n}=6,145) \\
{[}file-level{]} Median commits/month per AI agent config (file-age
normalized) & 0.4 & IQR: 0.2--1; median file age 4 mo
(\emph{n}=6,145) \\
{[}file-level{]} Median commits/month per CI/CD workflow (file-age
normalized) & 0.6 & IQR: 0.21--1.2; median file age 17 mo
(\emph{n}=24,436) \\
{[}cohort ≥2024-01-01{]} AI agent configs - median commits/month & 0.4 &
\emph{n}=6,144; median file age 4 mo \\
{[}cohort ≥2024-01-01{]} CI/CD workflows - median commits/month & 0.75 &
\emph{n}=15,852; median file age 8 mo \\
{[}file-level{]} CI/CD workflow files with ≤2 commits & 24.31\%
(5,940/24,436 files) & 23.77\% -- 24.85\% \\
Config files - exact duplicates across ≥2 repos (raw, incl.~forks) &
1,619 content groups; 5,733 file instances (18.75\%) & 18.31\% --
19.19\% of tracked config paths (\emph{n}≈30,581) \\
Config files - exact duplicates (fork-adjusted) & 425 content groups;
3,075 file instances (\textbf{10.06\%}) & excludes groups with
\textless2 non-fork repos; 653 fork repos in corpus (no separate CI
reported) \\
Config-file clone pairs (cross-org share) & 75.51\% cross-org
(66,839/88,517 pairs) & 21,678 within-org / 66,839 cross-org \\
Duplicate clusters spanning ≥2 organisations & 540 near/template; 702
exact & across 9,357 organisations \\
Repos with ≥1 credential pattern in configuration files & 3.18\%
(318/10,008; \textasciitilde1 in 31; lower-bound estimate) & 2.85\% --
3.54\%; regex-only; manual validation deferred (§9.3) \\
CI/CD workflow files declaring explicit permission scopes & 33.39\%
(8,158/24,436 files) & file-level coverage \\
{[}file-level{]} AI agent config files declaring explicit permission
scopes (indicative) & 31 true positives; 6 parser false positives
excluded & coverage 0.5\% (31/6,145); true coverage ≈ 0\% \\
{[}repo-level, supplementary{]} Repos with ≤1 commit to an AI agent
config file & 5.47\% (547/10,008) & 5.04\% -- 5.93\% \\
\end{longtable}
}

\textbf{Table 11. Version-control depth by file category.}

{\def\LTcaptype{none} 
\begin{longtable}[]{@{}lrlrr@{}}
\toprule\noalign{}
Category & Files & Median (IQR) & Exactly 1 commit & ≤2 commits \\
\midrule\noalign{}
\endhead
\bottomrule\noalign{}
\endlastfoot
AI agent configs & 6,145 & 1 (1--3) & 58.45\% & 74.26\% \\
GitHub Actions workflows & 24,436 & 7 (3--20) & 14.07\% & 24.31\% \\
Copilot instructions & 261 & 2 (1--5) & 34.48\% & 54.02\% \\
Cursor rules & 443 & 1 (1--2) & 62.75\% & 76.30\% \\
Claude config & 5,441 & 1 (1--2) & 59.25\% & 75.06\% \\
\end{longtable}
}

\emph{Permission scope coverage by category (formerly Table 12): AI
agent configs 31 TP / 6,145 (0.5\%); GitHub Actions 8,158 / 24,436
(33.39\%); Claude config 28 TP / 5,441 (0.51\%); Cursor rules 2 / 443
(0.45\%); Copilot instructions 1 / 261 (0.38\%). Agent-config permission
figures are indicative (parser n=31 TP).}

\begin{center}\includegraphics[width=\linewidth,height=0.6\textheight,keepaspectratio]{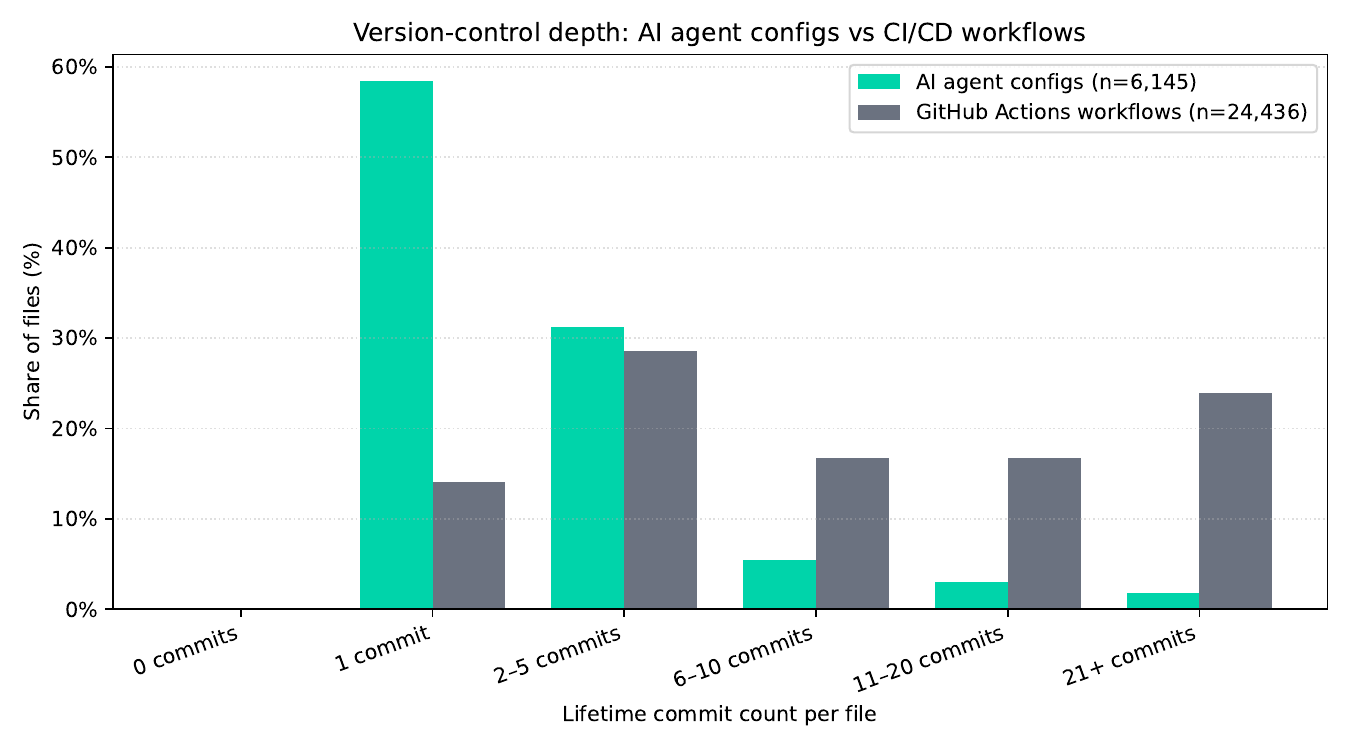}\end{center}

\emph{Figure 8. Raw lifetime version-control depth by file category. AI
agent configs concentrate at one commit (58.5\%); CI/CD workflows show
deeper history (median 7 commits). Age-normalised rates that bound the
artifact-age confound are in §7.2.1 / Figure 9. Source:
\texttt{report.json},
\texttt{paperStudy.versionControlDepthByCategory}.}

\paragraph{7.2.1 Age-normalized version-control
rate}\label{age-normalized-version-control-rate}

Raw commit counts confound file maturity: agent-configuration paths are
younger (median first-commit age \textbf{4} months for AI configs vs
\textbf{17} months for Actions; \textbf{66.5\%} of agent-config files
are younger than 6 months vs \textbf{22.5\%} for Actions). We therefore
report \textbf{commits per month} = lifetime commit count ÷ max(file age
in months, 1), measured from first commit to report generation.

\textbf{Table 11. Age-normalized version-control rate by file category.}

{\def\LTcaptype{none} 
\begin{longtable}[]{@{}
  >{\raggedright\arraybackslash}p{(\linewidth - 6\tabcolsep) * \real{0.2308}}
  >{\raggedleft\arraybackslash}p{(\linewidth - 6\tabcolsep) * \real{0.3077}}
  >{\raggedright\arraybackslash}p{(\linewidth - 6\tabcolsep) * \real{0.2308}}
  >{\raggedright\arraybackslash}p{(\linewidth - 6\tabcolsep) * \real{0.2308}}@{}}
\toprule\noalign{}
\begin{minipage}[b]{\linewidth}\raggedright
Category
\end{minipage} & \begin{minipage}[b]{\linewidth}\raggedleft
Files
\end{minipage} & \begin{minipage}[b]{\linewidth}\raggedright
Median file age (mo, IQR)
\end{minipage} & \begin{minipage}[b]{\linewidth}\raggedright
Median commits/month (IQR)
\end{minipage} \\
\midrule\noalign{}
\endhead
\bottomrule\noalign{}
\endlastfoot
AI agent configs & 6,145 & 4 (2--6) & 0.4 (0.2--1) \\
GitHub Actions workflows & 24,436 & 17 (6--40) & 0.6 (0.21--1.2) \\
Copilot instructions & 261 & - & - \\
Cursor rules & 443 & - & - \\
Claude config & 5,441 & - & - \\
\end{longtable}
}

\emph{Copilot, Cursor, and Claude rows omitted when \emph{n} is small
relative to aggregate claims; full category breakdown in
\citep{artifact}.}

\textbf{Cohort-matched subset (files first committed ≥ 2024-01-01)},
aligned with discovery filter \texttt{pushed:\textgreater{}2024-01-01}):
AI agent configs retain median \textbf{0.4} commits/month
(\emph{n}=6,144); Actions workflows \textbf{0.75} commits/month
(\emph{n}=15,852; median file age 8 months). The age confound explains
part of the raw depth gap (Figure 8) but does not eliminate it - Actions
workflows in the same repositories still revise more frequently per
month of existence.

\begin{center}\includegraphics[width=\linewidth,height=0.6\textheight,keepaspectratio]{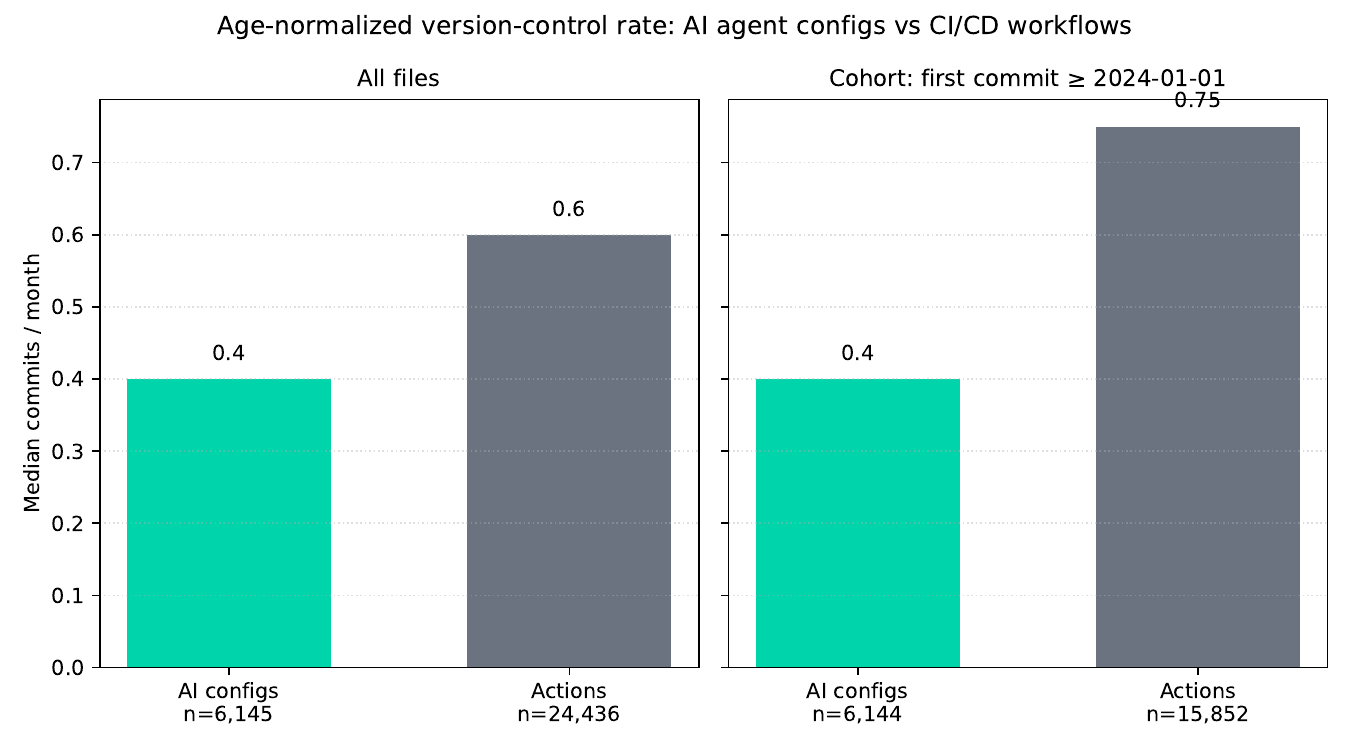}\end{center}

\emph{Figure 9. Median commits/month by category. Left panel: all files;
right panel: cohort with first commit ≥ 2024-01-01. Source:
\texttt{report.json}, \texttt{paperStudy.versionControlRateByCategory}
and \texttt{cohortMatchedVersionControl}.}

\textbf{Maintenance posture.} Among AI agent configs, \textbf{89.7\%}
are lightly maintained (1--5 commits) and only \textbf{10.24\%} actively
maintained (≥6 commits). GitHub Actions workflows in the same corpus are
\textbf{53.36\%} actively maintained - consistent with both raw depth
(Figure 8) and age-normalised rates (Figure 9).

\subsubsection{7.3 Interpretation}\label{interpretation}

§7 \textbf{confirms Gap 1 and partially supports Gap 3} (Table 1b)
through public-repository proxies. Gap 2 is a process-layer risk that
static config analysis cannot measure; it motivates §4.5 phase gates
independently of the prevalence study.

\textbf{Clone propagation} is the most SHA-256-robust headline for Gap
1: \textbf{10.1\%} of tracked configuration paths (fork-adjusted; 425
content groups) and \textbf{18.75\%} raw (1,619 groups) appear as exact
duplicates across repositories; \textbf{75.5\%} of clone pairs cross
organisational boundaries - agent configs circulate as undeclared shared
components, the failure mode SBOM requirements were designed to surface.

\textbf{Version-control immaturity} is also substantial: a 58\% majority
of agent configs are committed once and never revised (median = 1
commit); age-normalised rates remain lower than Actions in the same
corpus (§7.2.1), even after bounding the artifact-age confound. Nearly
90\% of agent configs fall into the lightly-maintained category versus
\textbf{53.4\%} actively maintained for Actions workflows.

\textbf{Permission and credential findings are indicative (Gap 3).}
Declared permission boundaries appear on fewer than \textbf{1\%} of
agent config files (31 true positives / 6,145; parser fragile) versus
\textbf{33\%} on Actions workflows - evidence that harness-specific
permission syntax is rarely used even when available. Credential
patterns appear in \textasciitilde1 in 31 repositories (lower-bound
regex estimate). These support the architectural emphasis on
deterministic pre-execution guardrails (§4.2, §4.3) rather than post-hoc
observability alone, but should not be weighted equally with
duplicate-propagation and depth metrics.

\subsubsection{7.4 Limitations of the
study}\label{limitations-of-the-study}

\begin{itemize}
\tightlist
\item
  \textbf{Sample size:} 10,008 repositories analysed, meeting the
  10,000-repository target (§9.3); adequacy is adequate.
\item
  \textbf{Fork metadata and duplicate rates:} Fork flags were backfilled
  for 8,988/10,008 completed repos (fork metadata unavailable for the
  remaining 1,020). Headline exact-duplicate prevalence uses a
  \textbf{fork-adjusted} rate (\textbf{10.1\%}; groups requiring ≥2
  non-fork repositories) with the \textbf{raw} rate (\textbf{18.75\%},
  including fork-network copies) reported alongside. Cross-organisation
  clone-pair rate (75.5\%) and cluster counts (702 exact clusters ≥2
  orgs) are computed on raw pairs and not fork-filtered. Residual fork
  ambiguity may remain where forks share content under different
  \texttt{owner/repo} names without a recorded parent link
  \citep{kalliamvakou}.
\item
  \textbf{Selection bias:} the corpus comprises public repositories
  discovered through configuration-file presence; findings characterise
  the ``already adopted AI agents'' population, not all of GitHub.
\item
  \textbf{Credential scanner:} regex-only detection; reported prevalence
  is a lower-bound estimate with possible false positives; manual
  validation sample deferred to §9.3.
\item
  \textbf{Artifact-age confound:} Agent-configuration file types are
  younger than GitHub Actions workflows (median first-commit age 4 vs 17
  months; §7.2.1). Raw commit-depth comparisons (Figure 8) understate
  Actions' maturity advantage. §7.2.1 reports commits-per-month and a
  cohort-matched subset (first commit ≥ 2024-01-01) to bound this
  confound; a residual rate gap remains after normalization.
\item
  \textbf{Permission parser:} agent-config formats lack standardised
  permission syntax; 6 of 37 raw matches (≈16\% FP rate on raw matches)
  were manually excluded; with only 31 true positives, a few additional
  FPs or missed TPs would materially shift the \textless1\% headline -
  treat as \textbf{indicative}.
\item
  \textbf{No causal claims:} prevalence does not establish that weak
  governance causes incidents, nor does it evaluate Rel(AI)Build
  deployment outcomes - the DRIFT vignette in Appendix D demonstrates
  detection behaviour only; full controlled developer outcomes remain
  §9.2.
\end{itemize}

\begin{center}\rule{0.5\linewidth}{0.5pt}\end{center}

\subsection{8. Limitations}\label{limitations}

We state the following architectural limitations explicitly to scope the
contributions of this paper. Empirical study limitations are in §7.4.

\textbf{Prompt injection is not solved.} The HITL gates and AC-scope
checks reduce, but do not eliminate, the risk of prompt injection. A
sufficiently well-crafted malicious instruction that mimics a legitimate
acceptance criterion may pass automated scope checks. The human review
gate remains the primary backstop.

\textbf{Local integrity, not adversarial signing.} The supply-chain
controls (SHA-256 integrity, HMAC lockfile, audit log) protect against
accidental corruption and automated tampering in a shared repository.
The default HMAC uses a machine-local key: any process running as the
same user can read the key and forge a valid stamp. These controls do
not defend against an adversary with direct, privileged access to the
developer's local machine, nor do they provide Sigstore-grade
cryptographic authentication without an organisational signing key
(future work).

\textbf{Cooperative trace linkage (determinism boundary).} Traceability
(§4.5) depends on the agent cooperatively invoking
\texttt{trace-update}; a prompt-injected or non-compliant agent can edit
files without registering AC linkage. The control plane detects such
gaps at the next spec-drift check but cannot prevent the edits - see
§4.5 trust boundary.

\textbf{Workspace trust boundary.} Integrity controls in §4.1 protect
\emph{this} project's installed agent definitions from tampering. They
do not protect against instructions embedded in a \emph{different},
untrusted workspace - e.g.~a cloned third-party repository whose
\texttt{AGENTS.md} or \texttt{.cursor/rules} attempt to steer an
operations agent toward \texttt{git\ config} manipulation (T7). Runtime
pre-exec hooks and \texttt{git-ops} hard constraints reduce but do not
eliminate this risk; developers should avoid running agents with
Bash/Git access inside unreviewed third-party workspaces.

\textbf{Does not validate semantic correctness of generated code.} The
architecture governs the process and configuration layer, not the LLM's
outputs. It cannot verify that generated code is logically correct,
functionally complete, or free of subtle bugs beyond what the testing
and review phases surface.

\textbf{Does not secure model-provider infrastructure.} Controls operate
at the harness and agent-definition layer. The confidentiality and
integrity of communications with LLM provider APIs (Anthropic, OpenAI,
Google) are outside this architecture's scope.

\textbf{Conformance testing focuses on enforcement properties, not
developer productivity outcomes.} The results in §6 demonstrate that the
deterministic mechanisms enforce their invariants as designed on
injected violations. Appendix D illustrates DRIFT detection on a single
governed build (N=1) without supporting effectiveness claims;
statistical inference on developer productivity outcomes requires the
controlled study in §9.2.

\textbf{Jaccard drift thresholds are operating defaults pending
calibration.} The 0.60/0.80 boundaries in §4.4 are not empirically
validated against behavioural-change metrics; sensitivity analysis is
deferred to §9.1.

\textbf{Blocklists require active maintenance.} The command and path
blocklists in §4.3 are grounded in known incidents and are therefore
bounded by what has been publicly documented. Novel attack variants,
including those targeting ecosystems not yet represented in Table 4, are
not covered until patterns are added.

\textbf{Runtime permission enforcement varies by substrate.} The control
plane guarantees that the canonical agent definition correctly declares
its permission tier and tool allowlist, and that this declaration is
faithfully translated into the target's native syntax. It does not
guarantee equivalent runtime enforcement across all harnesses. As noted
in §4.6, Claude Code enforces the tool allowlist at session start;
Cursor enforces access implicitly through scoping rules; and several
targets (Codex, Windsurf, OpenCode, Kiro) impose no tool-level runtime
enforcement. Organisations requiring uniform runtime enforcement should
audit their chosen harness's enforcement model independently.

\subsection{9. Future Work}\label{future-work}

While the architectural mechanisms (hashing, blocking, state machine
enforcement) are verified through conformance testing (§6) and the
motivating governance gap is quantified (§7), measuring the real-world
\emph{impact} of the control plane requires further empirical studies.

\subsubsection{9.1 Drift vs.~behaviour
correlation}\label{drift-vs.-behaviour-correlation}

Does a HIGH tokenisation drift (Jaccard \textless{} 0.60) reliably
predict a statistically significant change in agent task success rates?
A controlled experiment using SWE-bench \citep{sweagent} or a comparable
agent evaluation benchmark would apply measured Jaccard drift to the
system prompt and measure the resulting change in task-completion rate.
A monotonic relationship between drift bucket and performance delta
would validate the operational significance of the 0.60 and 0.80
thresholds.

\subsubsection{9.2 Developer study: overhead
vs.~benefit}\label{developer-study-overhead-vs.-benefit}

Appendix D provides a DRIFT-detection vignette on a single governed
build (N=1) and does not support comparative effectiveness claims. A
full controlled study requires randomised cohorts with equal N, a
standardised harness, pinned environments, and a bare-harness baseline
arm. Does the enforcement of AC-traceability and HITL gates measurably
reduce scope creep and reviewer-reported defects compared to that
baseline, and does the phase-gate overhead offset the time otherwise
lost to uncontrolled LLM recursion and rework? Outcome variables
include: time-to-merge, number of post-merge bug reports, and
self-reported cognitive overhead from the phase gates.

\subsubsection{9.3 Corpus completion}\label{corpus-completion}

Expand the prevalence corpus beyond the current \emph{N}=10,008
repositories (the 10,000-repository target is met) and publish the full
dataset, SQLite store, and companion reproduction scripts as arXiv
ancillary files \citep{artifact}. Re-apply fork-metadata enrichment on
the expanded corpus. Re-run near-clone detection at Jaccard thresholds
\{0.50, 0.60, 0.70, 0.80, 0.90\} to publish multi-threshold sensitivity
analysis. Conduct a manual validation sample (≥100 matches) on
credential-scanner hits to refine the 3.18\% exposure estimate.

\begin{center}\rule{0.5\linewidth}{0.5pt}\end{center}

\subsection{10. Conclusion}\label{conclusion}

LLM coding agents steer tools with write and execute authority, yet the
configuration and process layer that directs them lacks the governance
discipline routinely applied to application code and CI/CD pipelines.

A prevalence study of 10,008 public repositories (§7) confirms that this
thin governance surface is widespread. Its most robust finding is that
\textbf{10.1\%} of tracked configuration paths are exact duplicates
across independent repositories (fork-adjusted, SHA-256; \textbf{75.5\%}
of clone pairs cross organisational boundaries) - agent configs already
circulate as undeclared shared components. Agent configs are also
revised markedly less than CI/CD workflows in the same repositories (an
indicative finding that survives, but narrows under, age-normalisation).
Rel(AI)Build demonstrates that supply-chain integrity, pre-execution
permissions, attack-derived blocklists, and phase-gated traceability can
be enforced deterministically above any harness - conformance tests on
injected violations confirm each mechanism blocks its stated invariant
(§6).

What remains open is honest scope: cooperative trace linkage and prompt
injection are detectable but not fully preventable (§4.5, §8).
Controlled outcome studies on drift calibration, developer productivity,
and corpus expansion (§9.1--9.3) are required before claiming that these
controls improve developer outcomes at scale.

\begin{center}\rule{0.5\linewidth}{0.5pt}\end{center}

\subsection{Data and Artifact
Availability}\label{data-and-artifact-availability}

The full dataset, SQLite store, analysis pipeline, figures, and article
source are archived on \textbf{Zenodo} (DOI: 10.5281/zenodo.20780913)
\citep{artifact}; a subset is also attached as \textbf{arXiv ancillary
files} (\texttt{anc/}): \texttt{report.json},
\texttt{organizations.txt}, \texttt{prevalence-report-summary.json},
\texttt{ac-trace-table.csv}, \texttt{comparison-report-2-revised.pdf},
and the figure-reproduction scripts. The nine paper figures are vector
PDFs generated from these sources. Table 13 is self-contained; full
per-AC trace rows (Table S1), phase timelines, and attack-vector replay
notes are in the companion HTML/PDF report bundled in
\citep{impl-artifact}.

\begin{center}\rule{0.5\linewidth}{0.5pt}\end{center}

\subsection{Appendix A. Prevalence Study Reproduction
Parameters}\label{appendix-a.-prevalence-study-reproduction-parameters}

\textbf{Pipeline.} Custom corpus analysis pipeline: discovery → GitHub
REST API enrichment → clone detection → credential scan → permission
extraction → SQLite store → \texttt{report.json}. Reproduction scripts
and datasets: \citep{artifact}.

\textbf{Discovery seed list.} Built from GitHub Search API queries
stratified by star-count quartile:

{\def\LTcaptype{none} 
\begin{longtable}[]{@{}
  >{\raggedright\arraybackslash}p{(\linewidth - 4\tabcolsep) * \real{0.3333}}
  >{\raggedright\arraybackslash}p{(\linewidth - 4\tabcolsep) * \real{0.3333}}
  >{\raggedright\arraybackslash}p{(\linewidth - 4\tabcolsep) * \real{0.3333}}@{}}
\toprule\noalign{}
\begin{minipage}[b]{\linewidth}\raggedright
Quartile
\end{minipage} & \begin{minipage}[b]{\linewidth}\raggedright
Star range
\end{minipage} & \begin{minipage}[b]{\linewidth}\raggedright
Example query fragment
\end{minipage} \\
\midrule\noalign{}
\endhead
\bottomrule\noalign{}
\endlastfoot
Q1 & 0--10 &
\texttt{filename:CLAUDE.md\ stars:0..10\ pushed:\textgreater{}2024-01-01} \\
Q2 & 11--100 &
\texttt{filename:.cursorrules\ stars:11..100\ pushed:\textgreater{}2024-01-01} \\
Q3 & 101--1,000 &
\texttt{path:.github/workflows\ stars:101..1000\ pushed:\textgreater{}2024-01-01} \\
Q4 & \textgreater1,000 &
\texttt{filename:copilot-instructions.md\ stars:\textgreater{}1000\ pushed:\textgreater{}2024-01-01} \\
\end{longtable}
}

Query strings reflect the stratification design used to build the seed
list; exact per-query result counts are in the companion artifact
\citep{artifact}.

\textbf{Sampling frame.} Public GitHub repositories; English-primary
README not required; deduplication by \texttt{owner/repo} full name.
Fork metadata (\texttt{isFork}, \texttt{parentFullName}) backfilled via
GitHub REST API for 8,988/10,008 completed repos
(\texttt{enrich-fork-metadata}).

\textbf{Ingestion.} Analysed \emph{N}=10,008, meeting the
10,000-repository target (§9.3); 373 failures (3.59\%). Failure
categories (aggregate): GitHub API rate-limit backoff, repository
404/archived, enrichment timeout. Per-repo \texttt{failureLog} was empty
in the rendered report - failure characterisation is aggregate only.

\textbf{Clone detection.} Exact duplicates: SHA-256 content hash
(threshold-independent). \textbf{Fork-adjusted} exact duplicates:
content groups where ≥2 non-fork repositories share a hash (headline
\textbf{10.1\%}; 425 groups, 3,075 instances). \textbf{Raw} exact
duplicates (including fork-network copies): \textbf{18.75\%} (1,619
groups, 5,733 instances). Near/template duplicates: word-token Jaccard
with default threshold 0.80 (Table 5). MinHash/LSH candidate filtering
before pairwise Jaccard confirmation. Source: \texttt{report.json} →
\texttt{paperStudy.cloneAnalysis} (generated 2026-06-18).

\textbf{Parser validation.} Permission-scope parser: 37 raw matches on
agent configs, 6 manually confirmed false positives excluded (≈16\% FP
rate on raw matches; 31 true positives). Credential scanner: regex-only;
no systematic manual validation sample yet (deferred to §9.3).

\begin{center}\rule{0.5\linewidth}{0.5pt}\end{center}

\subsection{Appendix B. Governance-Preservation Compilation Argument
(Semi-Formal)}\label{appendix-b.-governance-preservation-compilation-argument-semi-formal}

This appendix states a \textbf{conditional semi-formal argument}, not a
mechanised proof. It closes the gap between citing formal workflow
methods \citep{aalst} and asserting compile-time governance
preservation.

\subsubsection{B.1 Definitions}\label{b.1-definitions}

\begin{itemize}
\tightlist
\item
  \textbf{D} - canonical agent definition (Markdown + YAML frontmatter).
\item
  \textbf{P} - governance property: (a) every tool in \texttt{tools(D)}
  belongs to the allowlist for \texttt{tier(D)}; (b) every path in
  \texttt{paths(D)} passes traversal defence; (c) the HMAC over the
  serialised lock record for \texttt{hash(D)} is valid.
\item
  \textbf{T(D)} - compiled target output for harness \emph{T} ∈
  \{Cursor, Claude Code, VS Code, \ldots\}.
\item
  \textbf{faithful\_encode(T, D)} - target \emph{T}'s parser recovers
  the same tool set and write-path set as declared in \emph{D} (syntax
  may differ; semantics under \emph{T}'s parser must match).
\end{itemize}

\subsubsection{B.2 Assumptions}\label{b.2-assumptions}

\begin{itemize}
\tightlist
\item
  \textbf{A1 (Pure compiler):} \texttt{compile(T,\ D)} is a pure
  function of \emph{D}; it does not add or remove tools or paths.
\item
  \textbf{A2 (Pre-compile gate):} Tier and traversal checks run on
  \emph{D} before any \texttt{T(D)} is written; violations abort
  compilation.
\item
  \textbf{A3 (Faithful encoding):} For each supported target \emph{T},
  \texttt{faithful\_encode(T,\ D)} holds for every \emph{D} that passes
  A2.
\end{itemize}

\subsubsection{B.3 Lemmas}\label{b.3-lemmas}

\textbf{Lemma 1 (Pre-compile certification).} If \emph{D} satisfies
\emph{P}, then A2 ensures (a) and (b) before compilation begins.

\textbf{Lemma 2 (Lock binds to canonical hash).} The HMAC stamp
references \texttt{hash(D)}, not \texttt{hash(T(D))}; lock validity is
independent of output format.

\subsubsection{B.4 Conditional property-preservation
argument}\label{b.4-conditional-property-preservation-argument}

If A1--A3 hold and \emph{D} satisfies \emph{P}, then for every supported
target \emph{T}, \emph{T(D)} satisfies \emph{P} under the interpretation
that (a)--(b) apply to the tool/path sets recovered by \emph{T}'s
parser, and (c) follows from Lemma 2.

\emph{Proof sketch.} By Lemma 1, (a)--(b) hold on \emph{D} before
compile. By A1, \texttt{tools(T(D))\ =\ tools(D)} and
\texttt{paths(T(D))\ =\ paths(D)} as sets. By A3, \emph{T}'s runtime
view matches those sets. By Lemma 2, (c) is unchanged by compilation. ∎

\subsubsection{B.5 Runtime caveat}\label{b.5-runtime-caveat}

Whether each harness \textbf{enforces} emitted permissions at runtime is
harness-specific (Table 2) and outside compile-time scope (§8). A4
(runtime enforcement equivalence) is explicitly \textbf{not} assumed.

Validating A3 (faithful encoding) per target - that the parser-recovered
tool and path sets match the canonical declaration - is deferred to the
expanded evaluation in §9.2; one-target parser-recovery tests constitute
a one-day exercise and are planned.

\begin{center}\rule{0.5\linewidth}{0.5pt}\end{center}

\subsection{Appendix C. Phase State Machine
(Summary)}\label{appendix-c.-phase-state-machine-summary}

The phase-gated lifecycle (§4.5, Figure 5) is implemented as a labeled
transition system over phases \{0, 1, 3, 4, 5, 7, 8\} with four HITL
pauses. \textbf{Table C1} lists the guard predicates enforced before
phase advancement; conformance tests in §6.2 exercise these guards with
injected violations.

\textbf{Table C1. Selected phase transition guards.}

{\def\LTcaptype{none} 
\begin{longtable}[]{@{}
  >{\raggedright\arraybackslash}p{(\linewidth - 4\tabcolsep) * \real{0.3333}}
  >{\raggedright\arraybackslash}p{(\linewidth - 4\tabcolsep) * \real{0.3333}}
  >{\raggedright\arraybackslash}p{(\linewidth - 4\tabcolsep) * \real{0.3333}}@{}}
\toprule\noalign{}
\begin{minipage}[b]{\linewidth}\raggedright
Transition
\end{minipage} & \begin{minipage}[b]{\linewidth}\raggedright
Guard (must hold)
\end{minipage} & \begin{minipage}[b]{\linewidth}\raggedright
Failure code
\end{minipage} \\
\midrule\noalign{}
\endhead
\bottomrule\noalign{}
\endlastfoot
\texttt{start\_phase(n)}, \emph{n} \textgreater{} 0 &
\texttt{end\_ts{[}n−1{]}} exists & \texttt{PHASE\ ORDER\ VIOLATION} \\
\texttt{end\_phase(n)} & \texttt{start\_ts{[}n{]}} exists &
\texttt{END\ WITHOUT\ START} \\
\texttt{end\_phase(n)} & required delegations ⊆ recorded{[}\emph{n}{]} &
\texttt{DELEGATION\ GATE\ FAILURE} \\
\texttt{end\_phase(0)}, tier ∈ \{M,L\} & \texttt{spec-reviewer} ∈
recorded{[}0{]} & \texttt{REVIEW\ GATE} \\
\texttt{end\_phase(n)} & if file-writing delegations occurred:
\texttt{securityScan{[}n{]}.status\ =\ clean} &
\texttt{SECURITY\ SCAN\ GATE} \\
\texttt{end\_phase(n)} & auto-fix iteration count ≤ 3 per phase
entry\footnote{Default cap is 3; overridable via
  \texttt{PACK\_MAX\_ITERATIONS} (range 1--5). Rationale: §4.5 lifecycle
  design.} & escalation to HITL \\
\end{longtable}
}

Phases 2 and 6 are reserved slots \footnote{Phase numbering (0, 1, 3, 4,
  5, 7, 8) is non-contiguous by design: phases 2 and 6 are reserved for
  future specialist sub-workflows without renumbering deployed systems.}.
Every transition appends to the hash-chained audit log. Full LTS
state/event definitions are in the reference implementation
(\texttt{workflow-state} core); behavioural equivalence is established
by conformance tests (§6.2), not formal verification.

\begin{center}\rule{0.5\linewidth}{0.5pt}\end{center}

\subsection{Appendix D. Illustrative DRIFT-Detection Walkthrough (N=1,
mechanism
vignette)}\label{appendix-d.-illustrative-drift-detection-walkthrough-n1-mechanism-vignette}

\begin{quote}
\textbf{Scope:} This appendix demonstrates Phase-8 spec-drift detection
on \textbf{one governed build} (N=1). It is \textbf{not} a comparative
evaluation; arm-level completeness, security replay, and test-count
metrics are \textbf{out of scope} and available only in the companion
artifact submitted as arXiv ancillary files \citep{impl-artifact}.
\end{quote}

We walked through a Task Management API specification - 26 acceptance
criteria (ACs), FastAPI / SQLAlchemy / Python 3.11+ - using the
Rel(AI)Build phase-gated lifecycle (\texttt{cursor-relAI} arm). A
CI-hardened comparator arm (\texttt{cursor-only}) was run in parallel
for internal calibration; its outcomes are \textbf{not} reported as
evaluation evidence here. This vignette does not replace §6 conformance
testing and does not support effectiveness claims at scale.

\subsubsection{D.1 Build context}\label{d.1-build-context}

The governed build scoped \textbf{CR-1} to core functional ACs;
cross-cutting ACs (security headers, CORS, token observability) were
tracked for follow-on CR-2. Both arms implement API-key authentication.
Full per-AC classification rows are in \citep{impl-artifact}, Table S1.

\subsubsection{D.2 DRIFT-detection
vignette}\label{d.2-drift-detection-vignette}

\textbf{Table 13. Phase-8 DRIFT detection (governed arm, n=26 ACs).}

{\def\LTcaptype{none} 
\begin{longtable}[]{@{}
  >{\raggedright\arraybackslash}p{(\linewidth - 2\tabcolsep) * \real{0.5000}}
  >{\raggedright\arraybackslash}p{(\linewidth - 2\tabcolsep) * \real{0.5000}}@{}}
\toprule\noalign{}
\begin{minipage}[b]{\linewidth}\raggedright
Event
\end{minipage} & \begin{minipage}[b]{\linewidth}\raggedright
Observation
\end{minipage} \\
\midrule\noalign{}
\endhead
\bottomrule\noalign{}
\endlastfoot
Phase-8 Guardian review & \textbf{DRIFT} on AC-018 - implementation
present, test trace missing \\
Merge blocked & Yes - Phase-8 exit withheld until trace gap resolved \\
Post-delivery testing audit & AC-018 test trace added; 26/26 ACs
verified \\
Dominant failure mode & \emph{Implementation without test trace}
(impl-only), not missing code \\
\end{longtable}
}

Phase-8 Guardian \textbf{DRIFT} surfaced AC-018 trace incompleteness
before merge - open-loop detection without guaranteed remediation within
the delivery window; the gap was closed in a post-delivery testing
audit. This illustrates the §4.5 traceability mechanism: the control
plane detects incomplete test linkage post-hoc; it does not guarantee
the agent registers tests during implementation.

\subsubsection{D.3 Mechanism
observations}\label{d.3-mechanism-observations}

\begin{itemize}
\tightlist
\item
  \textbf{Phase gates detect AC drift before merge} - DRIFT on AC-018
  blocked finalisation until the trace gap was closed.
\item
  \textbf{Impl-without-test is the dominant trace gap} - feature code
  existed for AC-018 but no verifying test was registered in the trace
  record at Phase-8 exit.
\item
  \textbf{Definition completeness bounds trace outcomes} - CR-2 scoping
  for cross-cutting ACs shows trace quality depends on canonical spec
  scope, not gate mechanics alone.
\end{itemize}

Agents optimise for scoped functional ACs; cross-cutting concerns
require explicit inclusion in the canonical change request or
CI-equivalent enforcement - motivating complete canonical definitions
and the §4.3 blocklist as a specification-independent baseline.

\subsubsection{D.4 Vignette scope
limits}\label{d.4-vignette-scope-limits}

\begin{itemize}
\tightlist
\item
  N=1; governed arm only in Table 13.
\item
  Single IDE harness (Cursor); GitHub Copilot is the AI model within
  Cursor, not the harness.
\item
  Comparator arm details, security replay scores, and test-count
  comparisons: \citep{impl-artifact} only - not paper claims.
\item
  CR-1/CR-2 phased scope means the full 26-AC spec was not completed in
  a single governed pass.
\end{itemize}

\begin{center}\rule{0.5\linewidth}{0.5pt}\end{center}

\emph{Incident sources (Table 4 blocklist / Table 6 threats, §5):
event-stream (2018), SolarWinds/SUNBURST (2020), Codecov (2021),
ua-parser-js (2021), xz-utils (CVE-2024-3094), Nx Console (2026).}

\bibliography{refs}

\end{document}